\documentclass[a4paper,fleqn,usenatbib]{mnras}

\usepackage[T1]{fontenc}
\usepackage{ae,aecompl}
\usepackage{appendix}
\usepackage{graphicx}	
\usepackage{amsmath}	
\usepackage{amssymb}	
\usepackage{booktabs}	
\usepackage{csquotes}	
\usepackage[normalem]{ulem}
\usepackage{url}

\usepackage{newtxtext,newtxmath}

\newcommand{\blu}{\textcolor{blue} }
\newcommand{\red}{\textcolor{red} }

\usepackage{tablefootnote,footnotehyper} 
\usepackage{array}
\newcolumntype{H}{>{\setbox0=\hbox\bgroup}c<{\egroup}@{}} 

\title[Population synthesis of $\delta$\,Sct stars]{Population synthesis of delta Scuti stars: the instability strip, period--luminosity relation, and large-separation distribution}

\author[Murphy et al.]{
Simon J. Murphy$^{1}$\thanks{E-mail: simon.murphy@unisq.edu.au} and
Anuj Gautam$^{1}$\thanks{E-mail: anuj.gautam@unisq.edu.au}
\\
$^{1}$ Centre for Astrophysics, University of Southern Queensland, Toowoomba, QLD 4350, Australia\\
}

\date{Accepted XXX. Received YYY; in original form ZZZ}
\pubyear{2026}

\begin{document}
\label{firstpage}
\pagerange{\pageref{firstpage}--\pageref{lastpage}}
\maketitle

\begin{abstract}
Interpreting the observed properties of intermediate-mass stars requires accounting for rotation, binarity, intrinsic population diversity, and measurement uncertainty.
We combine these effects in population-synthesis models of 1.4--2.5\,M$_{\odot}$ stars that include pulsation properties of $\delta$\,Sct variables. The synthetic populations successfully reproduce the distributions of stars observed in young associations, validating the mapping between intrinsic and observed stellar properties. Using these models, we infer a semi-empirical $\delta$\,Sct instability strip by matching synthetic and observed populations. The resulting instability strip is systematically hotter than widely used theoretical prescriptions, implying that $\delta$\,Sct stars are on average 0.1\,M$_{\odot}$ more massive than previously assumed, with 95\% of the population spanning approximately 1.50--2.30\,M$_{\odot}$. We then investigate the $\delta$\,Sct period–luminosity relation and reproduce its recently identified second ridge with mixture models in which roughly one-third of stars have the fifth or sixth radial overtone as their dominant mode; the fraction increases in younger populations. In contrast, unresolved binarity does not account for the second ridge. Finally, we predict the population-wide distribution of asteroseismic large separations, $\Delta\nu$, finding a peak between 6 and 7\,d$^{-1}$, in agreement with recent observations. We also demonstrate that $\Delta\nu$ is most reliably measured with the \'echelle technique, while autocorrelation methods can produce spurious detections when used in isolation. These results show that population synthesis provides a powerful framework for interpreting populations of $\delta$\,Sct stars and for linking stellar evolution, pulsation, and observation.
\end{abstract}




\section{Introduction}
\label{sec:intro}

Population synthesis models are common tools for investigating a wide range of astrophysical phenomena, from the formation of subdwarf B stars \citep{rodriguez-segoviaetal2025} to the impact of different stellar initial mass functions on the mass-to-light ratios of galaxies \citep{zonoozietal2025}. Binary population synthesis models are particularly prevalent and incorporate the effects of orbital evolution to describe the various phenomena that arise therefrom \citep[e.g.][]{rileyetal2022,zlietal2024,osbornetal2025}. Such models are useful for establishing the occurrence rates of rare events, or examining the interplay of complex astrophysical processes.

Simulating stellar populations with differing input physics can help to determine the expected properties of stars under those different conditions \citep{toonenetal2014}, and by comparing synthetic populations with observations, one can understand which variables are important \citep[e.g.][]{sharmaetal2017,emsenhuberetal2023}. With large sample sizes, it is also possible to identify trends from models that might not be apparent from individual observations, particularly when those observations have non-negligible uncertainties.

In this work, we explore a synthetic population of pulsating intermediate-mass stars (1.4--2.5\,M$_{\odot}$). Many stars in this mass range are delta Scuti variables, which pulsate primarily in pressure modes (p\:modes) driven by the $\kappa$-mechanism \citep{antocietal2019,kurtz2022}, with smaller contributions from other driving mechanisms \citep{houdek2000,antocietal2014,murphyetal2020d}. Low-order gravity modes (g\:modes) are also excited in $\delta$\,Sct stars \citep{breger2003}, and those with $n=-1$ and $n=-2$ are found at frequencies just below the fundamental radial mode in stars near the ZAMS \citep{gautametal2026}. Some $\delta$\,Sct stars also pulsate in high-order ($n\gtrsim30$) g\:modes, i.e. they are also gamma Doradus variables and hence $\delta$\,Sct--$\gamma$\,Dor hybrids \citep{hareteretal2010,grigahceneetal2010a}. Although $\delta$\,Sct stars can be found in the pre-main-sequence evolutionary stage \citep{zwintz2017,murphyetal2021a,steindletal2021a}, most of them are main-sequence (MS) stars. 

The $\delta$\,Sct instability strip is estimated by both theory \citep{dupretetal2004} and observation \citep{murphyetal2019} to be $\sim$2000-K wide, but it has long been apparent that not every star within the instability strip pulsates \citep{breger1969b}. For decades it was unclear whether all stars would be variable if only the observations were precise enough. When the Kepler Space Telescope was able to detect modes with only a few $\upmu$mag amplitudes, the requirement for precise photometry was satisfied and it did appear that there were non-pulsators inside the instability strip \citep{guziketal2014,guziketal2015}, although many of them could be explained away as being chemically peculiar or having uncertain temperatures and lying near the instability strip edges \citep{murphyetal2015a}. 
With {\it Gaia} came more precise stellar parameters \citep{gaiacollaboration2018a,gaiacollaboration2023a}, and it appeared that not only was the $\delta$\,Sct instability strip impure, but the pulsator fraction varies as a function of temperature \citep{murphyetal2019} or colour \citep{manietal2025}, luminosity \citep{readetal2024}, age \citep{murphyetal2024,berryetal2025} and rotation \citep{gootkinetal2024,murphyetal2024}. The need for a large sample and the existence of observational uncertainty make the instability strip's boundaries and purity a pertinent topic to investigate via population synthesis.

Another topic well-suited to analysis with a synthetic population is the period--luminosity ($P$--$L$) relation, which has seen much recent discussion \citep{ziaalietal2019,jayasingheetal2020,poroetal2021,baracetal2022,garciahernandezetal2024,readetal2024,manietal2025}.
{\it Gaia} parallaxes and the availability of all-sky space-based photometry have proven transformational in delivering precise luminosities and periods, with attention now turning towards the second ridge on the $P$--$L$ diagram. The periods, luminosities, and temperatures from realistic synthetic populations complement the flurry of recent observational work, and may offer insights into the origin of the second ridge.

This work is an extension of our work to synthesise stellar populations to provide realistic uncertainties on mass and age for field stars (\citealt{murphyetal2026a}; Paper\:I, hereafter). In Paper\:I, we showed that intrinsic metallicity spread, realistic rotation distributions, and the existence of binary companions have a large impact on the positions of stars on the HR diagram. That is, such positions are not determined solely by mass and age. These systematic effects, combined with observational measurement uncertainty, will bias and blur the inference of masses and ages from isochrones. Similarly, we expect them to influence the pulsator fraction, inferred instability strip edges, and the period--luminosity relation. These pulsation properties are the focus of this second paper. 

This paper is organised as follows. In Sec.\,\ref{sec:synth} we summarise the population synthesis models, 
including the distributions from which the stellar masses, ages, and metallicities are drawn, as well as the effects of rotation, binarity, and random observational uncertainty on the observed stellar properties. In Sec.\,\ref{sec:cepher}, we verify the accuracy of our synthetic population by comparing its stellar properties to observed stars in associations. We de-bias observations to infer the intrinsic boundaries of the $\delta$\,Sct instability strip in Sec.\,\ref{sec:IS}, and we derive a period--luminosity--temperature relation that we compare with observations in Sec.\,\ref{sec:PL}. The latter two topics are given longer introductions in their corresponding sections. Our final application is the prediction of the overall distribution of asteroseismic large separations, $\Delta\nu$, that one can expect to observe for $\delta$\,Sct stars (Sec.\,\ref{sec:overall}). Our conclusions are given in Sec.\,\ref{sec:conclusions}, and we provide an appendix where we compare the \'echelle method with the autocorrelation method for determining $\Delta\nu$ (App.\,\ref{app:autocorr}).

\section{Population synthesis}
\label{sec:synth}

We used the synthesised models from Paper\,I and we did not compute any new models or change any physics, here. Those models can be summarised as follows.

The underlying models are those from \citet{gautametal2026}. From these, we took models whose masses span 1.4--2.5\,M$_{\odot}$, and we resampled them according to the Salpeter IMF, ensuring that low-mass stars are more common. Their evolutionary tracks are sampled every 1\,Myr, independent of mass. Their metallicities were sampled following a Gaussian distribution around the solar value, specifically [Fe/H] = $-0.022\pm0.115$\,dex following \citet{willettetal2023}.

Simulated stars had a 70\% chance of being in an unresolved binary system, corresponding to inferred multiplicity fractions for A stars \citep{moe&distefano2017}. The primary stars were given rotational velocities from observed distributions \citep{zorec&royer2012}, and depending on whether they were binaries or not: instead of the fast-rotating distribution used for single stars, the primaries of binary systems were drawn from a bimodal distribution that includes a slowly-rotating component, representing tidal braking. However, we emphasise that no evolutionary effects on rotation were directly applied. 

The centrifugal effect of rotation was then applied to the temperature and luminosity, according to an inclination angle drawn from an isotropic distribution. Centrifugal effects were also applied to the stellar density ($\rho$), and mode frequencies were adjusted from their non-rotating counterparts in accordance with $\sqrt{\rho}$ \citep{ulrich1986}. Only radial modes were calculated; the density reduction dwarfs the first- and second-order effects of rotation on these modes.\footnote{For radial modes, the first order ($m\Omega$) terms are zero, because $\ell=m=0$.}

The companion masses were drawn from known mass-ratio distributions for A-type stars \citep{murphyetal2018}, and their luminosities calculated via a mass-luminosity relation \citep{ekeretal2015}. The luminosities of each component were added and their temperatures were a luminosity-weighted average. We applied no modification to pulsation modes based on whether a star had a binary companion. The final observables were perturbed by a Gaussian of 250\,K and 0.07\,dex in $\log L$ to simulate the effect of random observational uncertainty. These stellar properties are given the label `obs' in Paper\,I, and are the ones we use in this work unless otherwise indicated.

The stellar $T_{\rm eff}$ and $L$ are available from Paper I at various stages of the above processing, as summarised in Table\:\ref{tab:popsynth_data}.

\begin{table}
    \centering
    \caption{Stages in the construction of population synthesis models, with their abbreviated names, full names, and descriptions.}
    \begin{tabular}{lll}
    \toprule
        Abbrev. & Full name & Description \\
    \midrule
       `orig'  & {\it original} & original non-rotating models \\
       `mod'  & {\it model} & rotational effects added \\
       `inc'  & {\it inclined} & viewing angle effect added \\
       `bin'  & {\it binary} & binary effects added \\
       `obs'  & {\it observed} & observational uncertainty added \\
    \bottomrule
    \end{tabular}
    \label{tab:popsynth_data}
\end{table}

\section{Simulating clusters: Cep--Her}
\label{sec:cepher}

With the synthetic population, we aim to predict the properties of groups of stars. An obvious and useful application is to simulate stars in a cluster or association and attempt to reproduce their observed properties. This facilitates further inference. For instance, if asteroseismic properties of cluster members are available, it may allow an age to be determined for the ensemble.

Our accuracy in simulating clusters will generally decrease with age because the evolutionary effects of rotation gradually become more important, but this is not an issue for young clusters, such as the Pleiades or the recently-parametrised Cep--Her Complex \citep{kerretal2024}. In the following, we use the term `simulation' to distinguish a specific collection of models from the more generalised `synthetic population' above. 




The Cep--Her Complex consists of at least four main stellar associations, each of which consists of multiple subgroups with ages spanning approximately 25--100\,Myr \citep{kerretal2024}. It offers a unique opportunity to evaluate our population synthesis models because it contains 195\,$\delta$\,Sct stars, of which 126 form a clear group of stars near the ZAMS (the `ZAMS Group' in \citealt{murphyetal2024}; ``M24'' in this section hereafter). Furthermore, 70 $\delta$\,Sct stars in the ZAMS Group showed regular pulsations that allowed $\Delta\nu$ to be measured, which is much larger than for any other stellar population.

It has already been demonstrated with the Cep--Her Complex itself that for $\delta$\,Sct stars, $\Delta\nu$ depends on both age and rotation (M24), and this has been confirmed with stellar models \citep{gautametal2026}. This implies that to accurately reproduce the $\Delta\nu$ distribution, we must also be able to reproduce the rotation velocity distribution. In this example, we simulate the Complex using its known age and using the synthetic population. 

A potential hindrance is that A-type stars in Cep--Her have an excess of slow rotators (M24). We also aim to quantify that here.

\subsection{Data preparation}

We took the observed properties of the ZAMS Group from M24, specifically, the $\Delta\nu$, {\it Gaia}~{\tt vbroad}, and TIC $T_{\rm eff}$ data, and simulated the Cep--Her Complex as follows. 
\begin{enumerate}
    \item We applied age cuts to our synthetic population to include only stars between 25 and 100\,Myr, and $T_{\rm eff}$ cuts to the {\it observed} properties to match those observed for the Cep--Her ZAMS Group ($9500>T_{\rm eff}>7200$).
    \item We applied a $v\sin i$ lower limit to the synthesised values due to the characteristics of {\it Gaia}~{\tt vbroad} measurements. The Gaia RVS spectrograph has instrumental broadening of 26\,km\,s$^{-1}$ \citep{cropperetal2018,frematetal2023}, however, some intermediate-mass stars with lower {\tt vbroad} are reported. In the Cep--Her ZAMS Group, the lowest {\tt vbroad} is 18.3\,km\,s$^{-1}$, so as a compromise we adopted $v \sin i = 20$\,km\,s$^{-1}$ as our lower bound, corresponding also to the threshold above which \citet{frematetal2023} report that the {\tt vbroad} pipeline produces reliable values. We implemented this lower bound by selecting all simulated stars with $v \sin i < 20$, and setting their values to $v \sin i = 20$. This approach is more appropriate than dropping the slow rotators, since they still exist -- they are just observed to have higher {\tt vbroad}.  We refer to this sample as the `raw simulated' values for later reference.
    \item We then implemented a further selection function on the synthesised $v\sin i$ values, because M24 found that the ability to measure $\Delta\nu$ had strong dependence on {\it Gaia}~{\tt vbroad} (their figure 11). They were able to measure $\Delta\nu$ for 42 of the 88 stars (48\%) with {\it Gaia}~{\tt vbroad} $< 150$\,km\,s$^{-1}$, and for 10 of the 35 stars (29\%) with {\it Gaia}~{\tt vbroad} $> 150$\,km\,s$^{-1}$. We thinned our synthetic population by those percentages accordingly, so that the ratio of stars with measurable $\Delta\nu$ at different projected rotation velocities matched the observations. We also removed all stars in our synthetic population with $v\sin i>240$\,km\,s$^{-1}$ (2\% of the population), since none of the available {\it Gaia}~{\tt vbroad} values lay outside this range and it is unlikely that $\Delta\nu$ would have been measurable for them anyway.\footnote{M24 already showed that {\it Gaia}~{\tt vbroad} is a suitable substitute for $v \sin i$ for most applications, although they did note that {\tt vbroad} underestimates $v \sin i$ at slow velocities (50\,km\,s$^{-1}$) by 10-15\% -- a discrepancy too small to matter for $\Delta\nu$ in this application.} We call this the `modified simulated' distribution.
    \item We did not apply any luminosity cut to the synthesised population. M24 used absolute magnitude to constrain the ZAMS group to avoid contamination by some non-members of the Complex, but application of their constraints to the synthesised population did not remove a single star.
\end{enumerate}

\begin{figure}
\begin{center}
\includegraphics[width=0.48\textwidth]{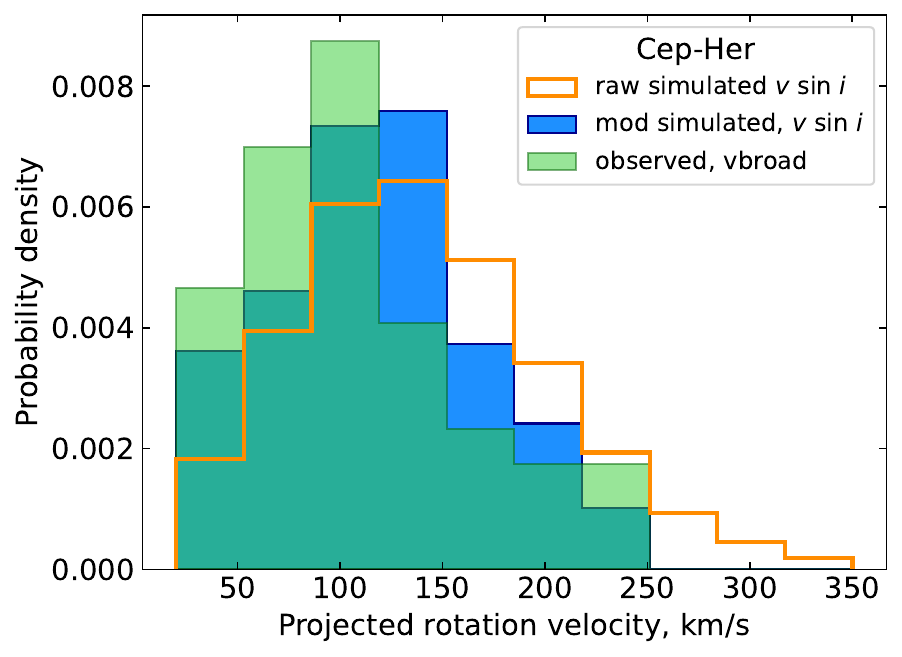}
\caption{The distribution of projected rotation velocities for observed $\delta$\,Sct stars in the Cep--Her Complex (green), and a synthetic population with similar properties (age, metallicity, and temperature range). The raw simulated distribution (orange line) removes only slow rotators, while the modified simulated distribution (solid blue) also removes the 2\% of very rapid rotators not seen in Cep--Her and accounts for $\Delta\nu$ measurability.}
\label{fig:cepher_rot}
\end{center}
\end{figure}

\subsection{Comparing rotation velocity distributions}
\label{ssec:rot_vel}

The resulting distribution of projected rotation velocities is shown in Fig.\,\ref{fig:cepher_rot}. The small tail of very rapid rotators ($v \sin i>240$\,km\,s$^{-1}$) is apparent in the raw simulated distribution, the absence of which in the observed distribution is not significant: applying a Poisson distribution and given a mean occurrence rate of 2\%, the probability of observing 0 very rapid rotators in a sample of only 70 stars is 0.247. The similarity of the observed and raw simulated distributions is remarkable: a two-sided Kolmogorov---Smirnov (KS) test gives a p-value of 0.053, indicating no significant difference between the distributions. This is strong validation of the population synthesis models.

We used probability density rather than raw counts in Fig.\,\ref{fig:cepher_rot} because the synthesised models greatly outnumber the observations, $m=13\,214$ to $n=70$, for an effective sample size of 69.6 (from $1/n_{\rm eff} = 1/n + 1/m$). Our approach is strongly preferable over randomly sampling 70 stars from the synthesised population, which yields $n_{\rm eff}=35$ and leads to various problems pertaining to small-number statistics. To illustrate this, we repeated those 70 draws in a Monte-Carlo simulation. In 1000 samples, we obtained p-values from two-sided KS tests that were $<0.05$ in 234 cases, and $<0.01$ in 55 overlapping cases, with a median p-value of 0.159. The Monte Carlo experiment shows that small-sample fluctuations can readily shift the outcome of a KS test across the significance threshold, demonstrating the advantage of comparing the observations with the full model distribution rather than down-sampling.

We now discuss the mystery of the excess of slow-rotators in Cep--Her compared to field stars, noted by M24: they compared the Cep--Her rotation velocities against a literature distribution calculated for {\it single} stars, whereas our simulated population also includes a slowly-rotating component to account for binary stars. Clearly the latter is required. This result suggests that Cep--Her contains many binary stars that are not readily identified via cursory analyses, such as searches for eclipses in TESS light curves or evaluating {\it Gaia} {\tt ruwe} values. In Fig.\,\ref{fig:cepher_rot} there remains a small shift towards lower {\tt vbroad} in the observations than in the modified simulated $v \sin i$ distribution, but with only $\sim$30 observed stars across the leftmost three bins, it is difficult to determine whether it is significant. If it is, possible explanations are: (i) that {\tt vbroad} is systematically underestimated at low values (see footnote), which would explain the leftmost bin only; or (ii) that the angular momenta and hence rotation velocities of stars born in large complexes differ from those of field stars upon which the model distribution is based.

\subsection{Comparing $\Delta\nu$ distributions}

\begin{figure}
\begin{center}
\includegraphics[width=0.48\textwidth]{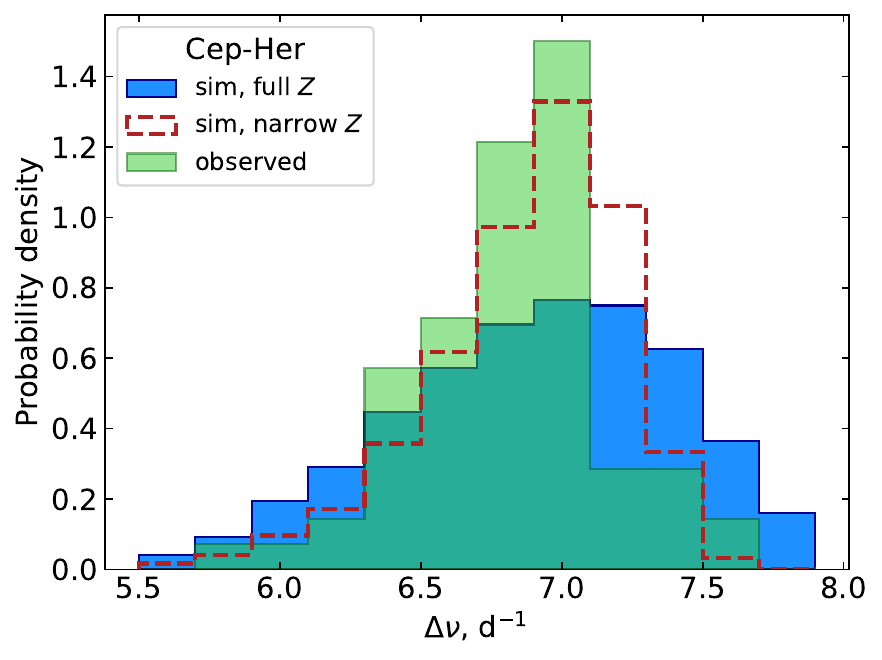}
\caption{The distribution of $\Delta\nu$ values for observed $\delta$\,Sct stars in the Cep--Her Complex (green), and a synthetic population with similar properties (such as age and temperature range; blue). The dashed red line shows a simulated distribution with a narrower range of metal mass fraction, $Z$.}
\label{fig:cepher_dnu}
\end{center}
\end{figure}

The pulsation properties of stellar populations are often summarised with their asteroseismic indices $\nu_{\rm max}$ and $\Delta\nu$. The former, the frequency of maximum power, cannot be studied with our models because they do not account for driving and damping. But the latter can. We analysed $\Delta\nu$ distributions of the observed and simulated populations, shown in Fig.\,\ref{fig:cepher_dnu}. The simulated distribution uses the modified simulated $v \sin i$ distribution and is coloured accordingly. The observed distribution is more narrowly peaked than the simulated one: the standard deviations are $\sigma_{\rm obs} = 0.33$ and $\sigma_{\rm sim} = 0.54$\,d$^{-1}$, respectively. The means also differ, with $\mu_{\rm obs} = 6.828$ and $\mu_{\rm sim} = 6.926$\,d$^{-1}$, respectively. Welch's $t$-test, which does not assume that the samples have equal variance, gives a $p$-value of 0.018, indicating that the means are significantly different. We now discuss possible causes for this.

We first considered that the age distribution of stars in the Cep--Her Complex is actually not uniform. Although the ages span 25--100\,Myr, 483 of the 958 potential members (50\%) considered by M24 were in subgroups spanning a much smaller age range of 25--43\,Myr, while only 54 of 958 (5.6\%) were in the oldest subgroup (80--100\,Myr). This heterogeneity was not accounted for in the simulation, but if it were, it would narrow the standard deviation and increase the mean of the simulated $\Delta\nu$ distribution further. We also considered that rejuvenation by mass transfer in binaries on the above timescales is improbable, and that the $\Delta\nu$ distribution would therefore not be materially affected.

The shape of the observed $\Delta\nu$ distribution is not Gaussian, and seems to be missing its higher-$\Delta\nu$ tail. This could point to an observational bias as a possible cause of the mismatch. Prior to the Cep--Her analysis, rather few $\delta$\,Sct stars in clusters had measured $\Delta\nu$ values, most of which lie in the $\sim$120-Myr-old Pleiades Cluster and have $\Delta\nu$ between 6.82 and 6.99\,d$^{-1}$ (\citealt{murphyetal2022}, see also \citealt{beddingetal2023}). When measuring $\Delta\nu$ for the Cep--Her stars, M24 made \'echelle diagrams with initial values of $\Delta\nu=6.9$\,d$^{-1}$ and searched either side of this value until they found a suitable $\Delta\nu$ (making vertical ridges in the \'echelles) and then manually optimised. This means that in `messy' pulsation spectra where there are multiple plausible values of $\Delta\nu$, it is more likely that values near the centre of the distribution would be chosen, potentially resulting in a narrowly peaked distribution. Although this method appears to be subjective, it meets the specific requirements of the present work in that it does yield $\Delta\nu$ values that match those of models, e.g. HD\,21434 with $\Delta\nu=6.94$ in Fornax--Horologium \citep{kerretal2022b} and TIC\,376872090 ($\Delta\nu=7.38$) in Cepheus Far North \citep{kerretal2022a}. Nonetheless, we experimented with autocorrelation techniques on the amplitude spectra of Cep--Her stars in an attempt to measure $\Delta\nu$ more objectively, but what we found was that autocorrelation techniques were not fit for purpose (Appendix.\:\ref{app:autocorr}).

A third possibility we considered is that the simulated population might have a broader metallicity distribution than the Cep--Her complex. When \citet{kerretal2024} analysed this complex, they used (fixed) solar-metallicity isochrones, while our synthetic population has a spread designed to match stars in the solar-neighbourhood. Of course, clusters and associations might have tighter values. We therefore repeated our analysis with a narrower range of metal mass fraction, keeping only models with $0.014 \leq Z \leq 0.016$ (solar is $Z=0.0142$). This distribution is shown as the dashed red line in Fig.\,\ref{fig:cepher_dnu}, and is a much better match to the observed values. A two-sided KS test gives a $p$-value of 0.105, indicating that the two distributions are not significantly different. The remaining (insignificant) difference between the observed and narrow-$Z$ distributions could be explained by the aforementioned observational bias, or minimised by fine-tuning the metallicity distribution. 

In summary, our simulation has reproduced the observed distributions of $\Delta\nu$ and projected rotation velocity of the Cep--Her Complex. The simulated $\Delta\nu$ distribution is broader than the observed one if the full metallicity range is used, but selecting a narrower range achieves a good fit. This suggests that population synthesis is capable of measuring asteroseismic ages and metallicities for clusters based purely on their $\Delta\nu$ distributions. It would be worth applying this method to other clusters if $\Delta\nu$ could be measured for many stars.

\section{Instability strip}
\label{sec:IS}

Here we evaluate whether population synthesis models can help to determine the boundaries of the $\delta$\,Sct instability strip, and whether the strip is intrinsically pure.

Several efforts have been made to ascertain the boundaries of the $\delta$\,Sct instability strip, with somewhat different outcomes. The most widely-used instability strip is the theoretical one computed with time-dependent convection (TDC) models by \citet{dupretetal2004}, who were the first to also include non-radial modes in the stability calculations. The resulting red edge was located between those determined by \citet{houdek2000} and \citet{xiong&deng2001}. As \citet{dupretetal2004} wrote, the calculation of the blue edge is well-determined \citep[e.g.][]{pamyatnykh2000} and the subject of less debate. By contrast, the red edge from the newer calculations by 
\citet{xiongetal2016} lies at a much cooler $T_{\rm eff}$, inconsistent with observations of the time \citep{uytterhoevenetal2011}, and this appears only slightly improved in recent iterations \citep{xiong2021}.

While the original TDC study presented only radial modes up to $n=4$ \citep{dupretetal2004}, this was soon expanded. \citet{dupretetal2005b} showed instability strip edges based on higher radial order modes up to $n=7$. Such high-order modes have since been observed and matched to pulsation models \citep{scuttetal2023}. Since \citet{dupretetal2005b} did not give the equations of those edges, we have measured and fitted straight lines to them using their figure 2 for later use. To four significant figures, for the \citet{dupretetal2005b} blue edge we measured
\begin{equation}
    \log L = -0.0008058 \times T_{\rm eff} + 
 8.205, \label{eq:dupret_blue}
\end{equation}
and for the red edge
\begin{equation}
    \log L = -0.001784 \times T_{\rm eff} + 
 13.01. \label{eq:dupret_red}
\end{equation}

In the TDC models, there is a free parameter describing the mixing length, $\alpha_{\rm MLT}$, whose value affects the location of the red edge. One generally constrains $\alpha_{\rm MLT}$ by comparison with observations \citep{houdek&dupret2015}. \citet{bowman&kurtz2018} suggested some mass-dependence in the value of $\alpha_{\rm MLT}$ to explain the simultaneous existence of high- and low-radial-order modes in observed $\delta$\,Sct stars, but none has so far been implemented. (Other works have preferred additional driving mechanisms like turbulent pressure to explain this observation, e.g. \citealt{antocietal2014}.) Instead, one commonly uses a fixed value of $\alpha_{\rm MLT}=1.8$, which \citet{dupretetal2004} found to match the 636 known $\delta$\,Sct stars of the time \citep{rodriguezetal2000}. Later, \citet{dupretetal2005b} showed that a lower $\alpha_{\rm MLT}$ would shift the instability strip redward, but observations were still too few and/or uncertain to optimise $\alpha_{\rm MLT}$ to the observed instability strip edges.

With \textit{Kepler} data, it became apparent that dozens of $\delta$\,Sct stars could be found at temperatures several hundred Kelvin hotter than the blue edge \citep{bowman&kurtz2018,murphyetal2019}, and some $\delta$\,Sct stars also existed at temperatures cooler than the red edge, albeit fewer in number. This raised a problem with the way that model instability strips were matched to observations: there were outliers beyond both edges, but they existed amongst a sea of {\it non-pulsators}, especially at the red edge. 
\citet{murphyetal2019} therefore suggested that the observed instability strip boundaries should be drawn with this in mind, and opted for edges at the locations where the pulsator fraction drops to 20\%. We use these edges in this work, and refer to them as the `20\% lines'.

In this work, we expand on a principle from Paper\:I, using our synthetic population to investigate whether biased stellar parameters might influence the mapping between the theoretical and observed instability strips. 
Suppose every star whose intrinsic ({\it model})\footnote{The italicised descriptors correspond to specific processing steps in the population synthesis as described in Paper\:I. The {\it model} properties are the global properties of the star after rotational (centrifugal) effects are applied, while the {\it observed} properties are those after applying all effects.} properties place it inside the instability strip were to pulsate and suppose one calibrated the theoretical instability strip against the extrinsic ({\it observed}) properties of those stars, namely, with the effects of inclination angle, binary companions, and measurement uncertainty added. Might that calibration, e.g. via the choice of $\alpha_{\rm MLT}$, be adversely affected by the mapping between intrinsic and extrinsic properties?

\begin{figure*}
\begin{center}
\hspace{0.01\textwidth}
\includegraphics[width=0.48\textwidth]{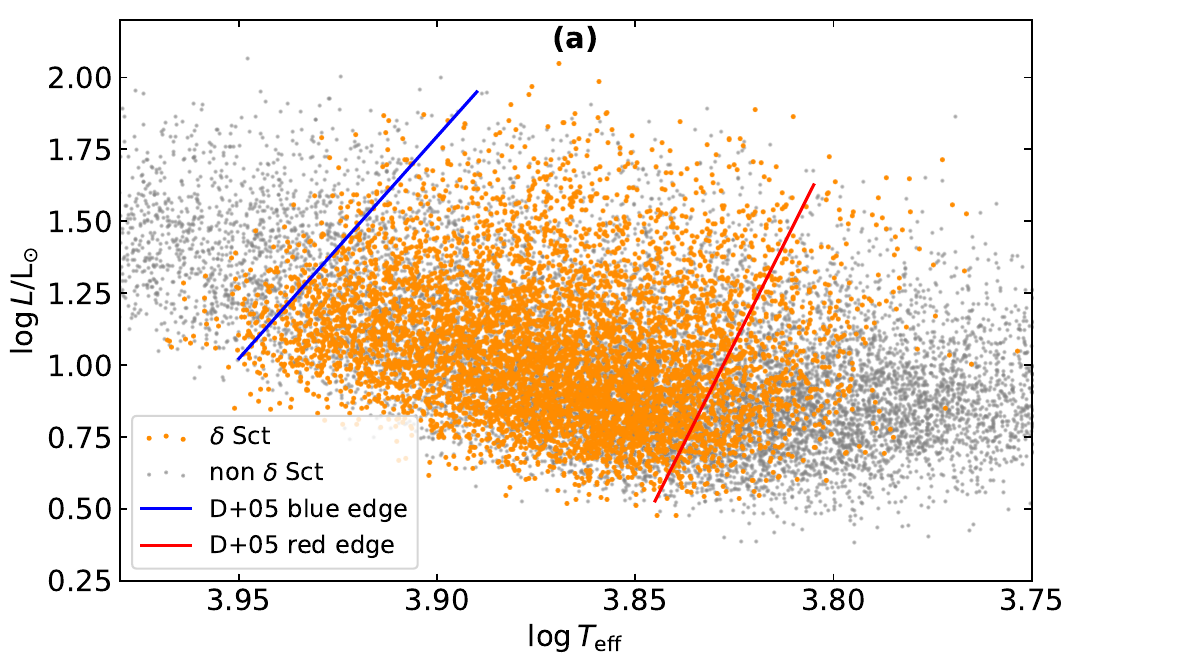}
\hspace{0.01\textwidth}
\includegraphics[width=0.48\textwidth]{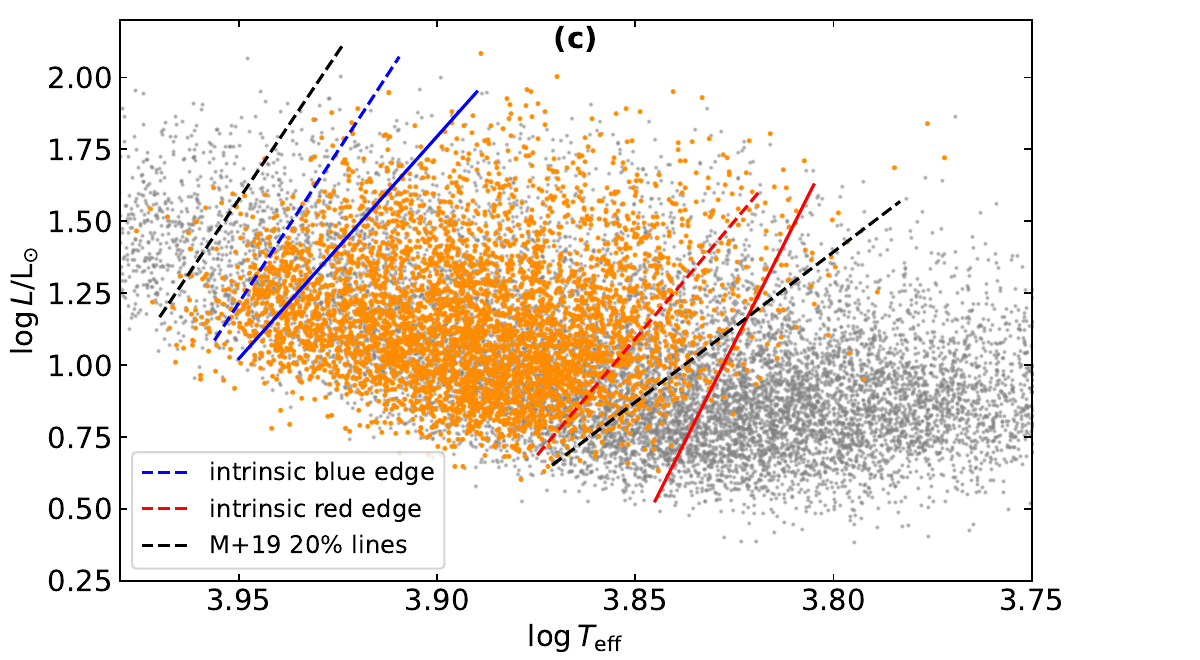}
\\
\includegraphics[width=0.49\textwidth]{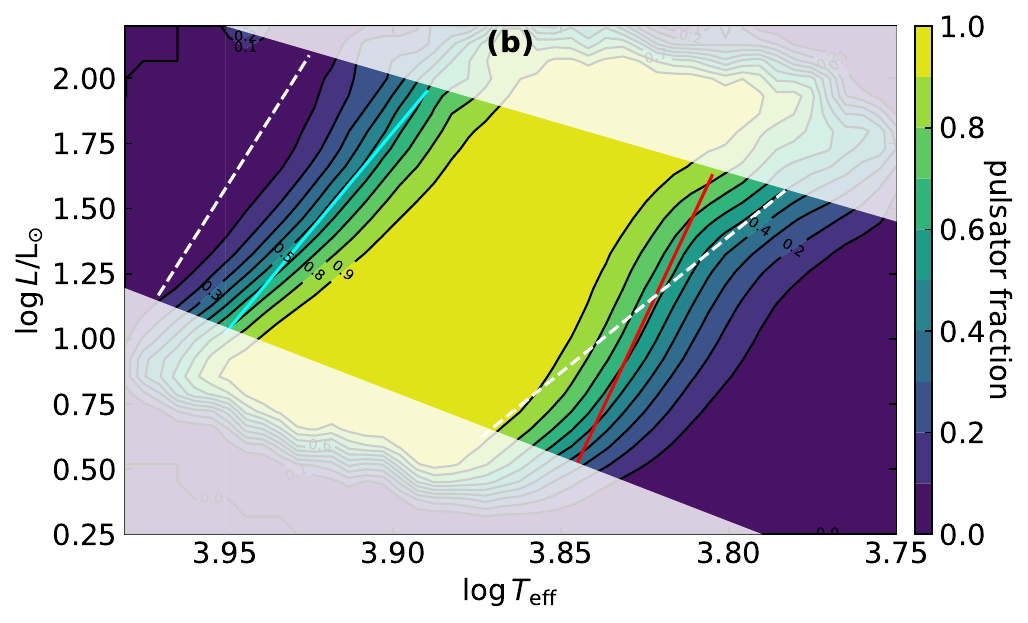}
\includegraphics[width=0.49\textwidth]{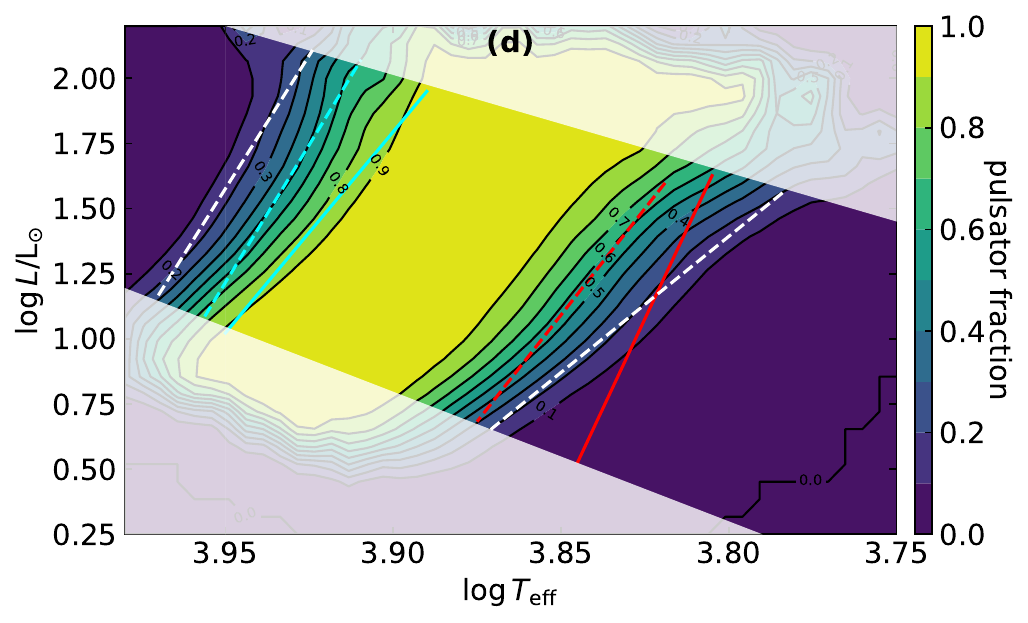}\\
\caption{{\bf (a)} The {\it observed} position of pulsators (orange circles) and non-pulsators (grey dots), if all stars whose {\it model} parameters place them inside the \citet{dupretetal2005b} instability strip (solid red/blue lines) are pulsating, and all stars outside are not pulsating. 
{\bf (b)} The contour plot of pulsator fraction corresponding to panel a, with the \citet{murphyetal2019} `20\% lines' added as dashed white lines. Areas with low star counts are greyed out. The line colour of the blue edge has been changed to cyan for visibility.
{\bf (c)} As panel a, but with a new intrinsic instability strip (dashed coloured lines), which produces an observed pulsator fraction that matches the `20\% lines' from \citet{murphyetal2019} (dashed black lines).
{\bf (d)} The contour plot of pulsator fraction corresponding to panel c, with the intrinsic edges (dashed red/cyan lines) and the \citet{murphyetal2019} `20\% lines' (dashed white lines) added. Notice that the 20\% lines now align with the contours corresponding to a pulsator fraction of 0.2.}
\label{fig:pure_strip}
\end{center}
\end{figure*}

Fig.\,\ref{fig:pure_strip}a shows the {\it observed} positions of simulated $\delta$\,Sct stars whose {\it model} properties place them inside the \citet{dupretetal2005b} instability strip (eqs\:\ref{eq:dupret_blue} and \ref{eq:dupret_red}). The stars are thinned by a factor of 100 for clarity. The mapping from {\it model} to {\it observed} properties causes a general shift to the right, hence many $\delta$\,Sct stars move beyond the red edge and many non-pulsators move into the instability strip from regions hotter than the blue edge. The pulsator fraction, being the ratio of these distributions, is shown as a contour plot in Fig.\,\ref{fig:pure_strip}b. It was calculated without thinning by gridding the data into 50 bins in the x-direction and 30 bins in the y-direction, corresponding to bin sizes of 0.005 in $\log T_{\rm eff}$ and 0.065 in $\log L$, then smoothing the result with a {\tt scipy} {\tt gaussian\_filter} with standard deviation of $\sigma=0.85$\,bins. Lowering the standard deviation does not fundamentally alter the location of the contours, but they become more ragged. The `20\% line' from \citet{murphyetal2019} is also shown for reference. It is apparent that the mapping between {\it model} and {\it observed} parameters has indeed moved the instability strip.

We attempted to reverse the above mapping by proposing a new semi-empirical (`intrinsic') instability strip, designed to reproduce the `20\% lines' once the extrinsic effects take place (Fig.\,\ref{fig:pure_strip}c). This involved changing the gradient and intercept of the edges by trial and error until the resulting pulsator-fraction contours match the 20\% lines (Fig.\,\ref{fig:pure_strip}d). Both of our new edges are substantially hotter, and the instability strip is 10\% narrower (at $\log L/{\rm L}_{\odot} = 1.1$). The equation of the intrinsic blue edge is
\begin{equation} 
\blu{\rm blue:} \log L/{\rm L}_{\odot} = -0.001071 \times T_{\rm eff} + 10.77
\end{equation}
or, an approximate form in $\log T_{\rm eff}$:
\begin{equation} 
\blu{\rm blue:} \log L/{\rm L}_{\odot} = -21.101  \times \log T_{\rm eff} + 84.57.
\end{equation}
The equation of the intrinsic red edge is
\begin{equation} 
\red{\rm red:} \log L/{\rm L}_{\odot} = -0.001009 \times T_{\rm eff} + 8.252
\end{equation}
 or, an approximate form in $\log T_{\rm eff}$:
\begin{equation} 
\red{\rm red:} \log L/{\rm L}_{\odot} = -16.528  \times \log T_{\rm eff} + 64.72.
\end{equation}
That the new edges are almost parallel in $T_{\rm eff}$ is coincidental.

There are two results from this analysis. 
The first concerns the peak pulsator fraction. Fig.\,\ref{fig:pure_strip} shows a central region where, after scattering effects, the pulsator fraction exceeds 90\%. Only in young associations have such pulsator fractions been observed \citep{murphyetal2024} (Berry et al., submitted); field stars typically show lower pulsator fractions peaking between 50 and 70\% \citep{murphyetal2019,gootkinetal2024,readetal2024,manietal2025}. Thus, 20--40\% of stars that intrinsically lie within the $\delta$\,Sct instability strip must be non-pulsators unable to be explained by scatter.
Mechanisms that cause this, namely the gravitational settling of helium out of the He\,{\sc ii} partial ionisation zone, are well known and have been reviewed elsewhere \citep[e.g.][]{turcotteetal2000,smalleyetal2017,durfeldt-pedrosetal2024}. The theory is corroborated by the aforementioned observations that show that the pulsator fraction increases with stellar rotation rate and decreases with stellar age. Intermediate pulsator fractions in the 120-Myr Pleiades cluster \citep[80\%;][]{beddingetal2023} are evidence that gravitational settling of helium takes a long time. Berry et al. (submitted) have recently analysed the decrease in pulsator fraction with age using various star clusters.

\begin{figure}
\begin{center}
\includegraphics[width=0.48\textwidth]{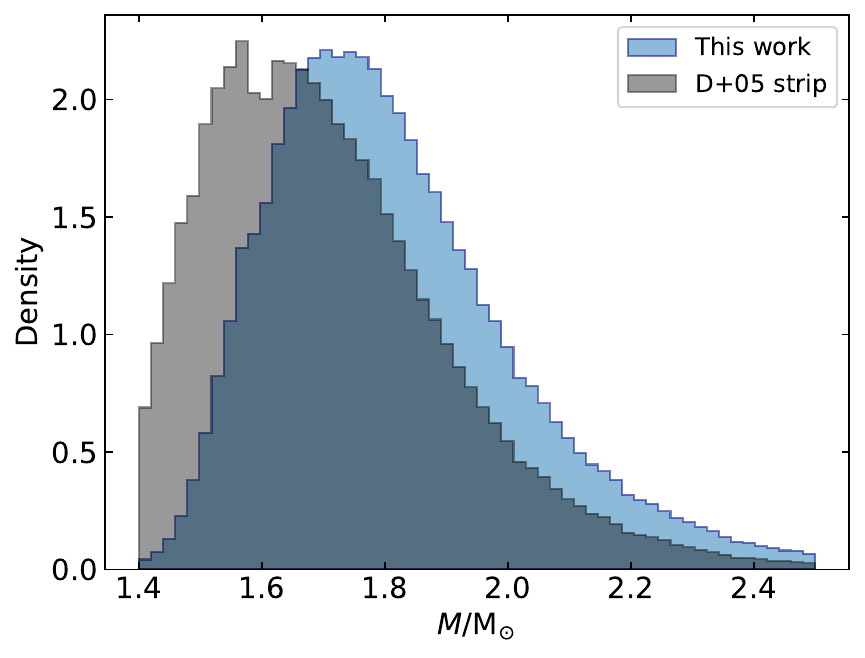}
\caption{The distribution of $\delta$\,Sct masses according to two different instability strips: the \citet{dupretetal2005b} instability strip (grey), versus the `intrinsic' instability strip presented in this work (blue). Both have been resampled with respect to their implied pulsator fractions in Fig.\,\ref{fig:pure_strip}.}
\label{fig:dsct_mass}
\end{center}
\end{figure}

The second result is that the average $\delta$\,Sct star is somewhat more massive than previously thought, at $\overline{M} = 1.81$\,M$_{\odot}$ rather than the 1.71\,M$_{\odot}$ implied by the \citet{dupretetal2005b} instability strip (Fig.\,\ref{fig:dsct_mass}), and correspondingly younger (566 vs. 657\,Myr, respectively). The 2$\sigma$ bounds on the mass range of $\delta$\,Sct stars are now 1.50 and 2.30\,M$_{\odot}$, to three significant figures. We arrived at these numbers after resampling the population of stars in each instability strip according to their respective pulsator fractions (in {\it observed}/extrinsic parameter space), which is necessary to down-weight the large number of low-mass stars that lie just inside the red edge of the instability strip (but where the pulsator {\it fraction} is small). There is a small uncertainty introduced by this process, because the observed pulsator fraction peaks at a lower value than our simulated ones, but since this occurs in the middle of the distribution, it has only a small effect on the above numbers ($<0.02$\,M$_{\odot}$ and <5\,Myr, respectively).
Existing models will need an increased $\alpha_{\rm MLT}$ to shift the instability strip to higher temperatures to account for this result.
There are also consequences for future modelling of the $\delta$\,Sct instability strip. 
If those models do not account for the various effects mentioned (rotation, viewing angle, binarity, and measurement uncertainty), then the intrinsic instability strip should be used for calibration instead of observed $\delta$\,Sct stars. Otherwise, we have provided the ingredients in Sec.\,\ref{sec:synth} and in Paper\:I for modellers to incorporate.

\section{Period--Luminosity relation}
\label{sec:PL}

The history of period--luminosity ($P$--$L$) relations for classical pulsators is long \citep{fernie1969}, and predates publication of the well-known relation for Cepheids \citep{leavitt&pickering1912}, with \citet{leavitt1907} observing that ``brighter variables have the longer periods''. A relation also exists for $\delta$\,Sct stars, with a shorter history underpinned initially by trigonometric parallaxes \citep{fernie1964} or by multicolour photometry and cluster memberships \citep{breger&bregman1975} and then by Hipparcos parallaxes \citep{mcnamara1997,mcnamara2011}. Since then, improvements have been made to the quality of the pulsation periods \citep{baracetal2022}, to the quality of the parallaxes \citep{ziaalietal2019,gaiacollaboration2023a}, and to the number of $\delta$\,Sct stars studied thanks to ASAS-SN and TESS light curves \citep{jayasingheetal2020,baracetal2022,readetal2024}. Various other samples of $\delta$\,Sct stars have seen their $P$--$L$ relations investigated, such as those in the Galactic bulge \citep{dekaetal2022}, the LMC \citep{martinezvazquezetal2022}, and the population II SX\,Phe variables \citep{cohen&sarajedini2012}, as reviewed by \citet{garciahernandezetal2024}. It appears that simultaneous application of $P$--$L$ relations in multiple colours may be particularly useful for constraining interstellar reddening \citep{guoetal2025}.

The $P$--$L$ relation arises because on the HR diagram the $\delta$\,Sct instability strip is quite narrow and iso-density contours are almost parallel to lines of constant luminosity for stars in hydrostatic equilibrium. Nonetheless, the existence of temperature (or colour) effects led some authors to adopt an additional colour term \citep{fernie1964,breger1969b,breger&bregman1975}. A colour term and even a metallicity term have been applied to the pulsations of chemically peculiar stars \citep{paunzenetal2002a} or metal-poor populations \citep{mcnamaraetal2007}, although \citet{liuetal2025} recently found metallicity to have a negligible impact on the $P$--$L$ relation of $\delta$\,Sct stars. Using our population synthesis models, we shall therefore consider a temperature term in this work but not a metallicity term. (We further justify this in Sec.\,\ref{ssec:z-term}.) We shall also construct period--density relations, expecting them to be tighter than $P$--$L$ relations.

Rotation, too, alters the $P$--$L$ diagram because of gravity darkening. \citet{garciahernandezetal2024} showed that the intercept of the $P$--$L$ relation changes when rotating stars are viewed pole-on versus equator-on. The shift can be so large that the fundamental-mode ridge of pole-on rapid rotators lies at higher luminosity than the first-overtone ridge of non-rotators. In practice, these two ridges blur together (e.g. \citealt{porettietal2008}), as we shall show here using models with a realistic distribution of rotation rates and inclinations.

The recent discovery of a second ridge in the $P$--$L$ diagram has sparked particular interest \citep{ziaalietal2019}. This second ridge lies roughly 0.3\,dex in $\log P$ to the left of the fundamental-mode ridge. There are many posited explanations. One is that the second ridge consists of binary stars, which cause the pulsator to appear over-luminous for its period, but \citet{ziaalietal2019} and \citet{baracetal2022} argued that the majority of stars on the second ridge are not binaries. Another explanation is that the second-ridge stars pulsate in higher-overtone modes, bearing resemblance to the multiple ridges seen in $P$--$L$ diagrams of red giants \citep{derekasetal2006}. \citet{poroetal2021} thus sought to categorise stars lying to the left of the fundamental-mode ridge into first-, second-, and third-overtone pulsators, following in the footsteps of \citet{breger&bregman1975}. A third explanation is that mode interactions, including amplitude modulation or resonant excitation, are responsible. Our models do not account for mode coupling, so we shall consider the first two hypotheses in our analysis of the $P$--$L$ relation via population synthesis, comparing against the \citet{baracetal2022} observations.

In this analysis, our simulation includes only stars that are in the intrinsic instability strip, and after resampling using the pulsator fraction as weights (see Sec.\,\ref{sec:IS}). All mode frequencies are adjusted for the centrifugal density reduction caused by rotation (Sec.\,\ref{sec:synth}), where rotating models are used.

\subsection{Simple case: non-rotating models}
\label{ssec:PL_nonrot}

\begin{figure}
\begin{center}
\includegraphics[width=0.48\textwidth]{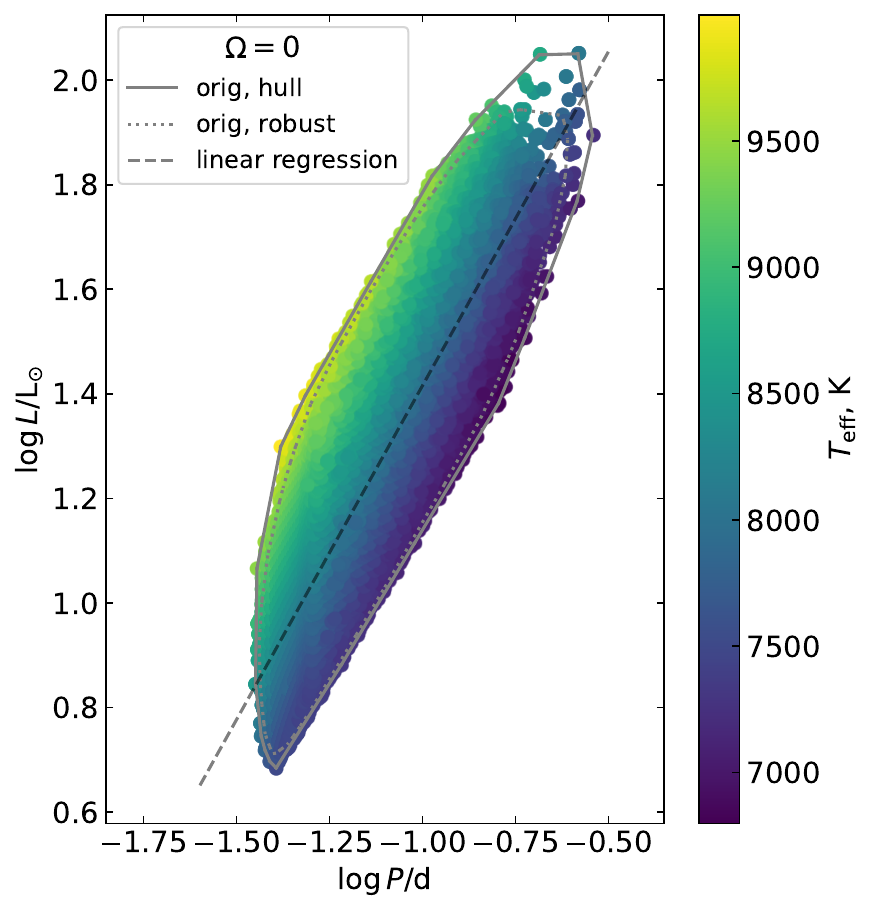}
\caption{The relationship between the period of the fundamental radial mode and the stellar luminosity for models of $\delta$\,Sct stars. Values used are the `original' values, without rotation, binarity, or observational uncertainty. Points are coloured by $T_{\rm eff}$. The calculated period--luminosity relation is shown as a dashed grey line. The solid and dotted grey lines are the convex hull and the robust convex hull of the data. The period range of the plot is extended to match the observational equivalent in \citep{baracetal2022}.}
\label{fig:PL}
\end{center}
\end{figure}

We begin by showing their $P$--$L$ relation in Fig.\,\ref{fig:PL}. The period is calculated for the fundamental radial mode (the $n=1$ p\:mode, abbreviated to `P1'). Here, we have used the {\it original} values for $\log L$ and $\log P$, i.e. before the addition of rotation, binarity, or observational uncertainty. The addition of these effects will be discussed shortly. Also shown in Fig.\,\ref{fig:PL} is the convex hull, which is the polygon with the smallest total area that encloses all the points. We will use convex hulls to compare samples with different physics, because they allow multiple samples to be overplotted in a way that is limited with scatter plots. We also define the robust convex hull, which is an outlier mitigation measure calculated by peeling layers of the hull (i.e. masking out hull vertices) until only some fraction of the data remain. We used the fraction 0.999.

We obtained the $P$--$L$ relation by linear regression
\begin{eqnarray}
{\it [original]} \quad\quad \log L/{\rm L}_{\odot} = 1.278 \log P + 2.694,\label{eq:PL_orig}
\end{eqnarray}
with an $R^2$ statistic of 0.724. Most of the remaining variance is due to $T_{\rm eff}$, with a little dependence on mass and almost no dependence on metal mass fraction. We also determined a $P$--$L$--$T_{\rm eff}$ relation via multiple linear regression
\begin{eqnarray}
    \log L/{\rm L}_{\odot} = 1.521 \log P + 4.790 \log T_{\rm eff} -15.752,\label{eq:PLT_orig}
\end{eqnarray}
where we used $\log T_{\rm eff}$ to keep the coefficients small, resulting in $R^2=0.998$. As a reminder, these are the values for non-rotating stars; significant departures are expected for rapid rotators and/or binary stars. 

There has been a recent surge in use of the period--density relation for $\delta$\,Sct stars, because it appears to give tighter ridges than a $P$--$L$ relation \citep{baracetal2022} (Mani et al submitted). The explanation is complex. It does not appear to be because density is easy to measure, since density can only be inferred by reference to models whilst also using $L$ as an input. In Paper I, we explained that isochrones can give biased masses and exhibit considerable scatter; even with population synthesis, we used the {\sc rapid} applet \citep{murphyetal2026a}, which allows users to get probabilistic physical stellar properties based on HR diagram position, to find in this work that density uncertainties can range from 20--50\%, depending on how well the luminosity has been measured, and how close the star is to the ZAMS (closer is better). Instead, the explanation lies in the benefit of an observed temperature constraint. Pairs of $T_{\rm eff}$--$L$ constrain density relatively well, and since mode frequencies scale as $\sqrt{\rho}$, density is a strong predictor once calibrated. By contrast, lines of constant luminosity on an H--R diagram span a range of temperatures, up to around $2000$\,K across the $\delta$\,Sct instability strip, as well as a range of masses (up to $\sim$0.8\,M$_{\odot}$), thus encapsulating a range of stellar densities and a corresponding range of mode frequencies.

\begin{figure}
\begin{center}
\includegraphics[width=0.48\textwidth]{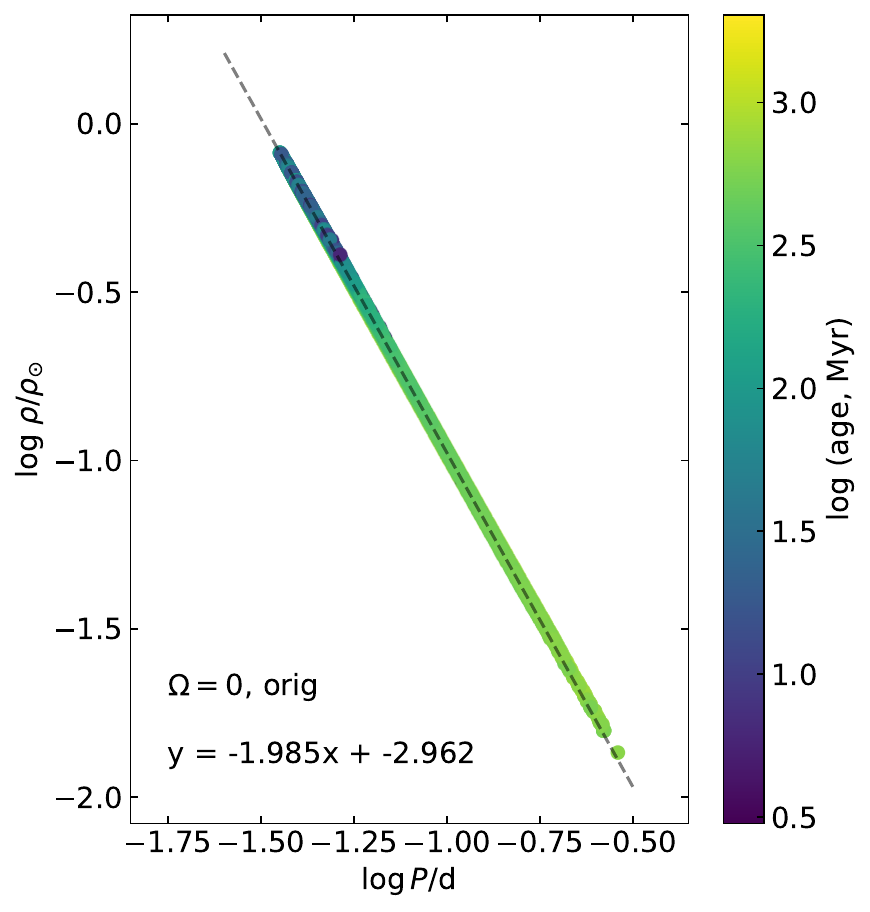}
\caption{The relationship between the period of the fundamental radial mode and the mean stellar density for models of $\delta$\,Sct stars. Although {\it original} values are used, rotation does not significantly alter this relation. Points are coloured by age, with pre-MS stars in the upper left and post-MS stars at the bottom right. The calculated $P$--$\rho$ relation is shown as a dashed grey line, whose equation is given.}
\label{fig:P-rho}
\end{center}
\end{figure}

Our models show that the $P$--$\rho$ relation is very tight (Fig.\,\ref{fig:P-rho}). Note that this does not change if we use the {\it mod} parameters and mode frequencies, since in our models these frequencies are calculated by scaling by $\sqrt{\rho}$. For a given mode, period and density are related by pulsation constants \citep[e.g.][]{cox1980},
\begin{eqnarray}
P = Q \left( \frac{\rho}{\rho_\odot} \right)^{-0.5},
\end{eqnarray}
which for the fundamental radial mode are approximately 0.033 \citep{fitch1981}, with some dependence on mass and metallicity \citep{dornan&lovekin2022}. This value is recoverable from the linear relation in Fig.\,\ref{fig:P-rho}, whose intercept is $2\log Q$.

While $P$--$\rho$ relations therefore have good utility if $\rho$ can be measured precisely, most of the literature concerns $P$--$L$ relations, so we shall continue our analysis with those.

\subsection{Advanced case: with rotation, binaries, and observational uncertainty}
\label{ssec:PL_rot}

\begin{figure}
\begin{center}
\includegraphics[width=0.48\textwidth]{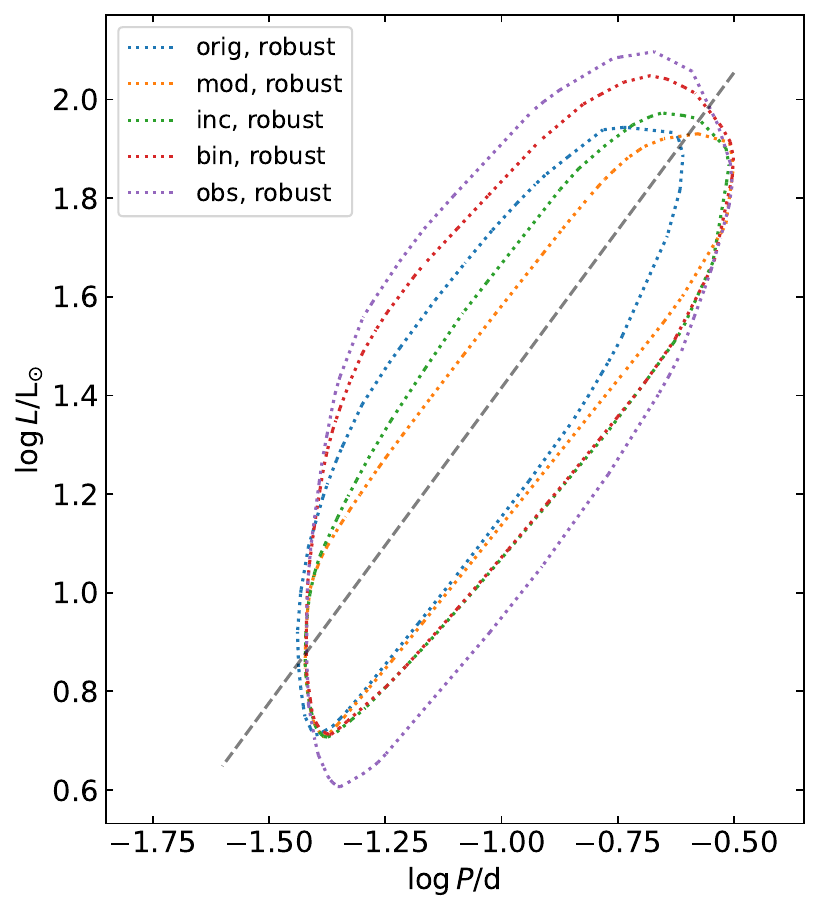}
\caption{Using robust hulls (defined Sec.\,\ref{ssec:PL_nonrot}) and with reference to the {\it original} hull from Fig.\,\ref{fig:PL}, we show the effect of adding rotation (`mod'), inclination effects (`inc'), binary effects (`bin'), and observational uncertainty (`obs') to models of the $P$--$L$ relation. The linear regression line from Fig.\,\ref{fig:PL} is shown as the dashed grey line.}
\label{fig:hulls}
\end{center}
\end{figure}

The addition of rotation shifts the distribution of models in the $P$--$L$ diagram to the right (Fig.\,\ref{fig:hulls}). This is because rotation decreases the mean stellar density and increases the mode periods in proportion to $\sqrt{\rho}$. Thus, the $P$--$L$ relation for the `mod' data whose robust hull is shown in Fig.\,\ref{fig:hulls} differs from that for the `orig' data (Eq.\,\ref{eq:PL_orig}). The inclusion of viewing angle effects substantially broadens the overall distribution of models. Binary effects leave the lower-right of the distribution untouched, but extend the upper end to higher luminosities, as can be expected. And finally, the inclusion of observational uncertainty extends this distribution in both directions in luminosity.

For a population with a realistic rotation velocity and inclination angle distribution, the gradient and intercept of the relation change to
\begin{eqnarray}
{\it [inclined]} \quad\quad \log L/{\rm L}_{\odot} = 1.176 \log P + 2.510, \label{eq:PL_inc}
\end{eqnarray}
or when a $T_{\rm eff}$ term is added,
\begin{eqnarray}
    \log L/{\rm L}_{\odot} = 1.524 \log P +4.844 \log T_{\rm eff}- 15.958. \label{eq:PLT_inc}
\end{eqnarray}
The above equations do not include binaries or observational uncertainty. They are perhaps the most useful forms, given that Gaia DR4 is imminent and it will then be much easier to detect and exclude binary systems. However, to demonstrate that the second ridge is not caused by binarity, we will need an equation that includes binary systems, which is
\begin{eqnarray}
{\it [binary]} \quad\quad \log L/{\rm L}_{\odot} = 1.162 \log P + 2.527. \label{eq:PL_bin}
\end{eqnarray}
Finally, for later use, we will need a $P$--$L$--$T_{\rm eff}$ relation based on the {\it observed} parameters, which is
\begin{eqnarray}
{\it [observed]} ~~ \log L/{\rm L}_{\odot} = 1.349 \log P + 2.573 \log T_{\rm eff} - 7.721. \label{eq:PLT_obs}
\end{eqnarray}

The full set of relations, with and without a $\log T_{\rm eff}$ term, for all five stages of parameters is given in Appendix\,\ref{app:PLRs}.

\subsection{Comparison with Barac et al.'s (2022) observations}
\label{ssec:PL_obs}
\begin{figure}
\begin{center}
\includegraphics[width=0.48\textwidth]{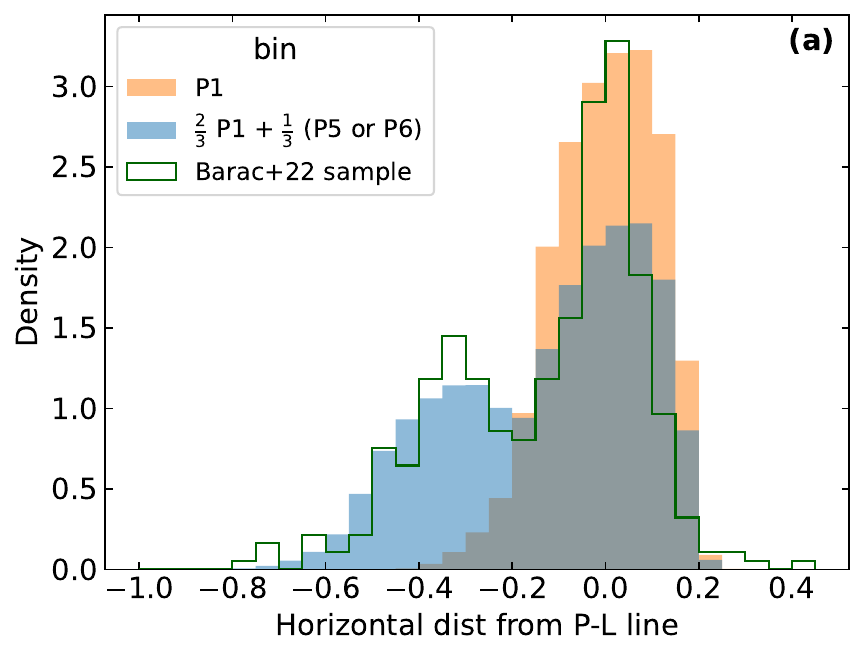}
\includegraphics[width=0.48\textwidth]{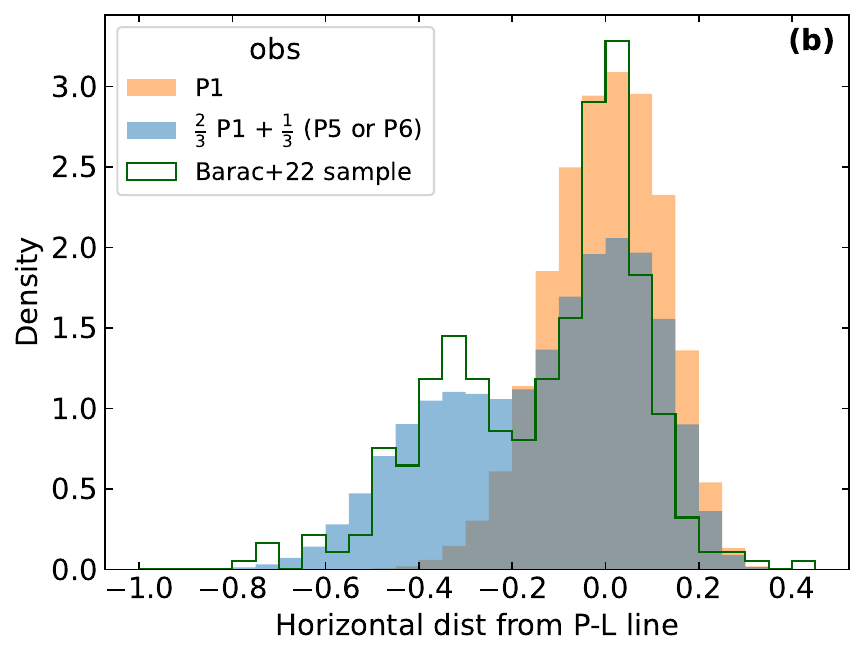}
\caption{{\bf (a)} Using the {\it binary} stellar properties, the horizontal distance of data points to the $P$--$L$ relation is shown under three different scenarios. In light orange, the distance of fundamental-mode (P1) pulsators from the relation in Eq.\,\ref{eq:PL_bin}. In blue, the mixture of fundamental mode and $n=5$ or $n=6$ (P5; P6) pulsators from the same relation. In dark-green, the observations published by \citet{baracetal2022} and their distance from their relation (their eq.\,4). {\bf (b)} As (a), but using the {\it observed} properties instead.}
\label{fig:horiz}
\end{center}
\end{figure}

It is common practice to parametrize the observed P--L relation in terms of horizontal distance from the fundamental-mode ridge (e.g. \citealt{ziaalietal2019}). This is particularly helpful since some relations are calculated in absolute magnitude (such as by \citealt{baracetal2022}), and some in luminosity (such as the models here). To calculate the horizontal distance, we used table\:2 from \citet{baracetal2022} along with their equation 4 to calculate the horizontal distance of the observed sample, and did the same for our models using Eq.\,\ref{eq:PL_bin}, without a $T_{\rm eff}$ term, to keep the models and observations on equal footing.
We plot both, along with an experiment to ascertain the nature of the second ridge, in Fig.\,\ref{fig:horiz}a, and we provide the same plot in {\it observed} properties for comparison (Fig.\,\ref{fig:horiz}b). We will now discuss these extensively, starting with the tail to the right of the fundamental ridge.

The P1 histogram of the models is constructed by setting every model to be a fundamental mode pulsator. We see in Fig.\,\ref{fig:horiz}a that it cannot explain the excess of observed stars with a horizontal distance of $x>0.20$, meaning that these stars cannot be explained as rapid rotators seen equator on (or similar). Since \citet{baracetal2022} were careful in examining the pulsation periods and even removing some RR\,Lyr variables, the observed excess is perhaps contributed by outliers in luminosity instead. When observational uncertainty is added (Fig.\,\ref{fig:horiz}b), the P1 histogram extends to higher $x$ values, but still cannot explain the highest observed $x$ values ($x>0.3$). Perhaps those stars are extreme outliers, or pulsate in some mode other than the fundamental mode. Note, though, that models with observational uncertainty do predict stars to exist in the range $x=0.1$--0.3, and these should not be discarded (cf. \citealt{jayasingheetal2020}).

The P1 histogram is also considerably broader than the observed fundamental ridge, and on both sides. If the effect had been one-sided, then a difference in the binary star fraction could be the explanation. Instead, the answer is probably rotational broadening. Since the \citet{baracetal2022} is based on a ground-based catalogue, it is likely that it is over-represented with high-amplitude pulsators (HADS), which in turn are over-represented with slow rotators \citep{breger2000}. Thus, rotation and inclination effects are less important for these observations than they are in the representative sample of stellar models.

We now turn our attention to the second ridge. The {\it binary} P1 histogram embeds binary population synthesis, yet cannot explain the second ridge by itself (the orange histograms in Fig.\,\ref{fig:horiz} have no second ridge). Hence, our models show conclusively that binarity is not the cause of the second ridge.

We attempted to evaluate the hypothesis that high-overtone pulsators or harmonics are responsible for the second ridge. We aimed to construct a mixture model of high-$n$ pulsators and P1 pulsators whose horizontal-distance histogram matched the observed one. The observed characteristics corresponding to the second ridge histogram are: (i) a height, in histogram density, that is about half the height of the P1 ridge; (ii) a location centred at $x\approx-0.33$; (iii) a gap between the two ridges that is about two histogram bins wide and is located at $x \approx -0.2$; and (iv) a termination between $x=-0.5$ and $-0.6$. 

We first experimented with a scenario wherein some fraction of stars have their strongest peak at exactly twice the frequency (half the period) of P1. Regardless of the chosen fraction, the second ridge was located too far to the right, hence, this explanation is inadequate. We repeated this experiment using the sum of the frequencies of P1 and P2 (the $n=2$ radial mode), which gave marked improvement, but a second ridge that was slightly too far to the left, giving a gap between it and the main ridge that was too large. By trial and error, we found a qualitatively good match for a mixture model in which 70\% of stars are P1 pulsators and the remaining 30\% are evenly split between having their highest peak at 2P1 and at P1+P2. We call this the `combinations and harmonics mixture model'. Since our models do not incorporate any information on the ratio of driving and damping, we do not argue for this model on any {\it physical} basis. We remark only that it comes close to matching the observations.

We also ran different experiments in which stars pulsated in other radial modes. Again by trial and error, we found a satisfactory match with mixture models in which one-third of stars pulsate in either the $n=5$ or $n=6$ mode instead of the fundamental mode (`the P5:6 mixture model'). This scenario is represented with the blue histogram in Fig.\,\ref{fig:horiz}. The P5:6 mixture model was a better visual match than one in which only P5 modes were used, or where the second-ridge stars were evenly split between P4, P5, and P6 pulsators. We prefer the P5:6 mixture model over the combinations and harmonics mixture model because several real $\delta$\,Sct stars have now been modelled whose strongest modes are at $n\geq5$ \citep[e.g.][]{kerretal2022a,murphyetal2023,scuttetal2023}. We chose not to optimise for the ratio of stars pulsating in the P5 or P6 modes vs. the fundamental mode because we know that there is significant $T_{\rm eff}$ and rotational variance in the relation, and because our models do not account for near-degeneracy effects such as mode-coupling (see e.g. \citealt{pamyatnykh2003}). Nonetheless, since stars dominated by P2, P3, and/or P4 modes were not required to obtain the match presented, near-degeneracy effects might be influencing the observed distribution (see discussion in \citealt{baracetal2022}). In particular, we notice that in 27\% of our models, the frequency of the combination P1+P2 is within 1\,d$^{-1}$ of the P5 mode frequency, although we have not applied second-order (or higher) frequency corrections due to rotation.

The one substantial difference between the P5:6 mixture model and the observations is again the broad fundamental ridge of the mixture model. This was common among all of our mixture models. It is because of this sample-specific effect that the $p$-value of a two-sided KS test is 0.023 ({\it binary} properties vs. observations), indicating that the two populations are significantly different.
Nonetheless, three of the four of the observed characteristics (ii--iv above) are well reproduced by the P5:6 mixture model. We conclude by noting that the additional broadening present in the {\it observed} properties histogram in Fig.\,\ref{fig:horiz}b produces a poorer match with observations: the second ridge is slightly less distinct. The implication is that faint $\delta$\,Sct stars from surveys such as ASAS-SN, OGLE, \textit{Kepler}, PLATO, or Roman, whose luminosities have larger uncertainties, would not be expected to show a clear gap between the main and second ridges.

\subsection{Comparison with other data sets}
\label{ssec:PL_obs_other}
Other works on the $\delta$\,Sct $P$--$L$ relation published after \citet{baracetal2022} have investigated larger numbers of stars \citep{readetal2024,garciahernandezetal2024}. We focus on \citet{readetal2024} because they had a simple selection function and they presented a histogram of horizontal distance from the $P$--$L$ relation (their figure 8). 

\begin{figure}
\begin{center}
\includegraphics[width=0.48\textwidth]{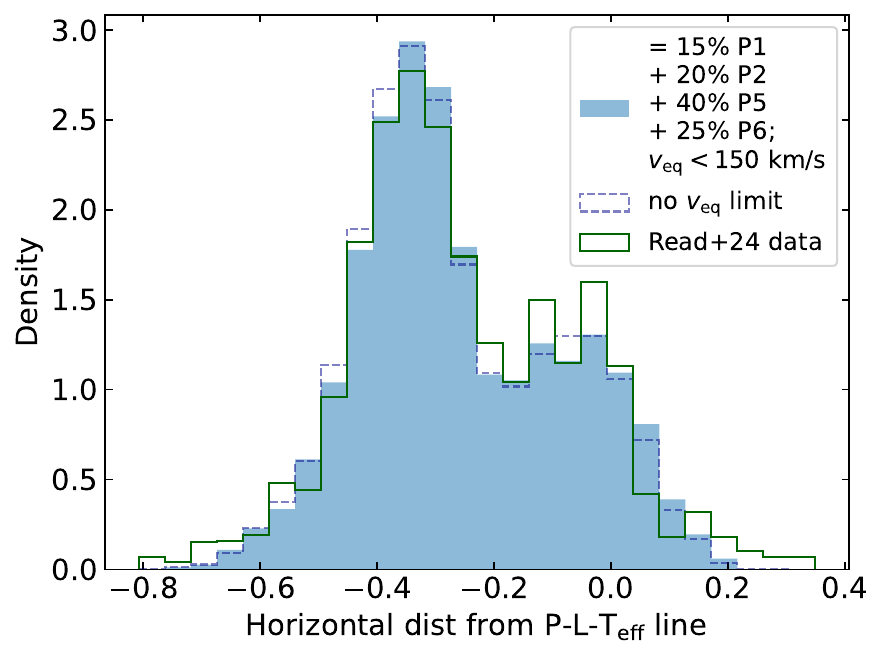}
\caption{A mixture model consisting of 15\% fundamental radial mode (P1) pulsators, 20\% first overtone mode (P2) pulsators, 40\% $n=5$ p-mode (P5) pulsators, and 25\% $n=6$ p-mode (P6) pulsators is a good match to the 850-star observed sample (green) from \citet{readetal2024}. The solid blue histogram shows the models when limited to $v_{\rm eq} < 150$\,km\,s$^{-1}$, while the dashed purple line has no $v_{\rm eq}$ limit. A $P$--$L$--$T_{\rm eff}$ relation (Eq.\,\ref{eq:PLT_obs}) for the {\it observed} properties was used in both cases.}
\label{fig:PLT_read_hist}
\end{center}
\end{figure}

\citet{readetal2024} specifically looked at a narrow vertical strip in the colour-magnitude diagram, such that temperature effects should be negligible. This presents an opportunity to apply our $P$--$L$--$T_{\rm eff}$ relation. Their sample was magnitude-limited, so unlike ground-based catalogues it is not biased towards luminous, high-amplitude, slowly rotating stars. They found that the second ridge became increasingly dominant towards higher (fainter) absolute magnitudes, and they also noted a smaller peak between the main and second ridges that they attributed to first-overtone ($n=2$) pulsators. 
We reproduce their histogram in Fig.\,\ref{fig:PLT_read_hist}. 

We constructed a mixture model consisting of 15\% fundamental radial mode (P1) pulsators, 20\% first overtone mode (P2) pulsators, 40\% $n=5$ p-mode (P5) and 25\% $n=6$ p-mode (P6) pulsators that yields a similar histogram using {\it observed} parameters and our corresponding $P$--$L$--$T_{\rm eff}$ relation (Eq.\,\ref{eq:PLT_obs}). We corroborate their extra peak at $x=-0.13$. To mimic \citet{readetal2024}, we used a 50-K slice starting at 8150\,K, which corresponds roughly to the middle of the observational instability strip. We note that a two-sided K--S test indicates no significant difference with a histogram constructed without this $T_{\rm eff}$ slice (p\:value = 0.93). We experimented with limiting the simulated sample to stars with $v_{\rm eq}<150$\,km\,s$^{-1}$ to mitigate extreme gravity darkening, which leaves 4148 models of the original 10\,802, and we plotted both in Fig.\,\ref{fig:PLT_read_hist}. The simulated sample reproduces the observed features and confirms the observed first overtone (P2) ridge, but this ridge is less distinctive for higher or absent $v_{\rm eq}$ limits. We expect that our luminosity scatter, at 0.07\,dex, is higher than that of the \citet{readetal2024} sample, and that this is causing more blurring of these ridges in the models than in the observations. (Our luminosity scatter was 2.6 times higher than the \citealt{baracetal2022} sample.) Thus, we expect that a sample constructed with smaller observational uncertainty but the full rotation velocity distribution would still reproduce the P2 ridge.

More complicated mixture models are again possible, e.g. where the probability of pulsating in higher-order modes depends on the stellar temperature and luminosity, but our simple one appears adequate. Our models do not account for driving and damping, so the mixture models of different modes are allocated randomly, i.e. not as a function of $T_{\rm eff}$ of the model or some other property. Coincidentally, \citet{tarczaynehezetal2026} recently found that $n=5$--6 modes are also driven in Cepheid variables.

Other data sets could be modelled in the same way. It is important to note that sample selection affects the histogram density of the fundamental and higher-order ridges. This can be seen in Figures\:\ref{fig:horiz} and \ref{fig:PLT_read_hist}. The bright sample of \citet{manietal2025} lies intermediate to these samples, in the sense that the fundamental and higher-order ridges have roughly the same heights in the histogram. We have not tried to reproduce the latter distribution here, because its bright stars ($G<7$ mag) have much smaller luminosity uncertainties than we have budgeted for. This sample would be worth exploring in the future, since very near to the ZAMS, populations are expected to contain fewer rapid rotators and fewer binaries (Paper I), enabling a purer characterisation of the $\delta$\,Sct pulsators. The larger but heterogeneous samples collated by \citet{garciahernandezetal2024} would be challenging to model because the constituent stars do not belong to a single, well-defined stellar population, but it is reassuring that the authors found the $P$--$L$ relations of those samples to be mutually consistent.

\subsection{Should there be a metallicity term?}
\label{ssec:z-term}

Our sample has only a narrow metallicity distribution compatible with stars in the solar neighbourhood. Nonetheless, our setup allows us to evaluate whether a metallicity term could be a useful addition to a $P$--$L$--$T_{\rm eff}$ relation.

In addition to the intrinsic spread, we added Gaussian-distributed measurement uncertainty of scale 0.15\,dex to [Fe/H], typical for a spectroscopic metallicity measurement for a field A-type star. We then compared the quality of a $P$--$L$--$T_{\rm eff}$ fit with a $P$--$L$--$T_{\rm eff}$--$Z$ fit for our 366\,921 models using their {\it observed} parameters. The adjusted $R^2$ value remained constant at 0.703, but the log-likelihood showed improvement from $2.4039\times10^5$ to $2.4079\times10^5$ and via the rms error we calculated a modest 0.11\% reduction in scatter.

Since homogeneous spectroscopic metallicities are seldom available and the improvement to the relation is modest, we have presented our results without a metallicity term.

\section{Predicting the overall $\Delta\nu$ distribution}
\label{sec:overall}

\begin{figure}
\begin{center}
\includegraphics[width=0.48\textwidth]{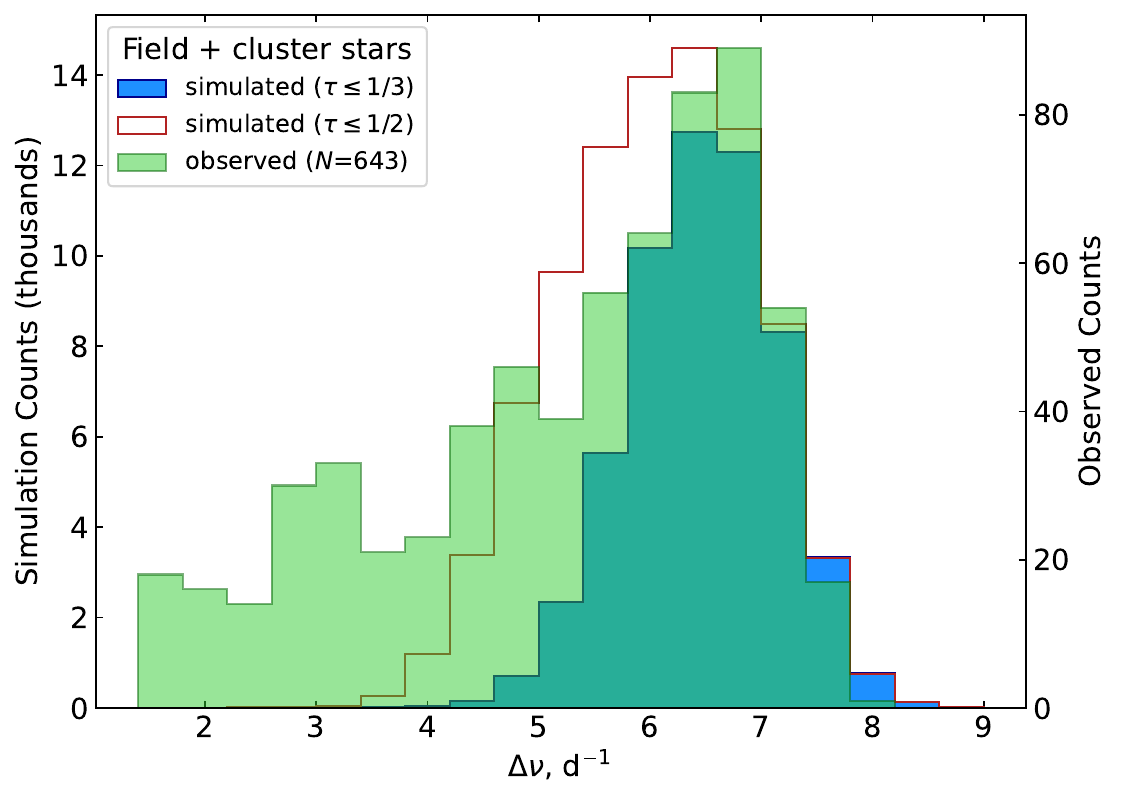}
\caption{The predicted $\Delta\nu$ distribution for $\delta$\,Sct stars younger than one-third ($\tau<1/3$) or one-half ($\tau<1/2$) of their expected MS lifetime are shown as the solid blue histogram and brown line histogram, respectively. The observed sample of 643 stars from \citet{boloukietal2026} is shown as the semi-transparent light-green histogram.}
\label{fig:field_dnu}
\end{center}
\end{figure}

By combining the previous applications, estimating the expected $\Delta\nu$ distribution of $\delta$\,Sct stars is trivial. Based on observations by \citet{beddingetal2020} and arguments by \citet{murphyetal2023}, we assume that regular patterns from which $\Delta\nu$ can be measured are only present in the first third of a star's MS lifetime. We measured the TAMS age as a function of mass across our synthetic population and simulated a young population whose ages are less than one-third of the TAMS age ($\tau<1/3$) for their mass and metallicity. 
We thinned the number of rapid rotators in this synthetic population exactly as we did for the Cep--Her simulation (Sec.\,\ref{sec:cepher}), to account for the fact that rapid rotators rarely show identifiable regularity in their pulsation patterns \citep{pandaetal2024a}. As with our $P$--$L$ relation analysis, we resampled according to pulsator fraction.
The resulting $\Delta\nu$ distribution is our prediction for field\footnote{The stars may belong to clusters or associations, but we do not assume that they do.} $\delta$\,Sct stars whose pulsation spectra exhibit regular patterns (blue histogram, Fig.\,\ref{fig:field_dnu}). The mean and standard deviation are 6.46 and 0.68\,d$^{-1}$, respectively.

During final preparation of this manuscript, \citet{boloukietal2026} published an observed $\Delta\nu$ distribution of 643 $\delta$\,Sct stars obtained via machine learning. We have added this to Fig.\,\ref{fig:field_dnu}. The agreement with our model prediction at the higher $\Delta\nu$ end is remarkable. This indicates that the real population is well understood: young $\delta$\,Sct stars are born with metallicities consistent with a solar-neighbourhood metallicity prior, and models accurately predict their $\Delta\nu$ values (see also \citealt{murphyetal2023,scuttetal2023}). This remains true with relatively crude treatment of the effects of rotation, as long as the underlying distribution of rotation velocities is correct (Paper\:I). Otherwise, the high $\Delta\nu$ end of the distribution would not be well reproduced.

The lower $\Delta\nu$ side of the distribution, below $\Delta\nu \approx 5.5$\,d$^{-1}$, is not as well reproduced. We considered that we might have prematurely discarded $\delta$\,Sct stars for losing their regular patterns at $\tau=1/3$, so we extended the analysis to $\tau=1/2$ (brown line in Fig.\,\ref{fig:field_dnu}). This has the effect of over-predicting the number of stars with $\Delta\nu \approx 6$ and under-predicting the number with $\Delta\nu<4$\,d$^{-1}$. We note that for ages of $\tau\geq1/2$, avoided crossings affect even high-order p\:modes that contribute to $\Delta\nu$ measurements \citep{gautametal2026}. In addition, at these older ages, the pulsator fraction of $\delta$\,Sct stars drops off (Berry et al., submitted). Both phenomena imply that regular patterns will become rare at $\tau>1/2$. We therefore suspect that the machine-learning detections of $\Delta\nu<4$\,d$^{-1}$ might be spurious. In particular, if some of the excess of observed patterns near $\Delta\nu=3$\,d$^{-1}$ were in fact half of the true values -- as is known to occur, see discussion in Appendix\:\ref{app:autocorr} -- then these would fill in the gap between the $\tau<1/2$ histogram and the observed histogram at $\Delta\nu=6$. Finally, regarding the left-most observations at $\Delta\nu\leq2$, our simulation shows such values to be achieved in only the most rapid rotators ($\overline{v_{\rm eq}} > 200$\,km\,s$^{-1}$), and to be biased towards high-mass stars ($\overline{M} > 2.2$\,M$_{\odot}$) with higher-than-average metal mass fractions. Of these, it is the high rotation rate that is most problematic, since it is rare to see regular patterns in such stars \citep{murphyetal2024}.

\section{Conclusions}
\label{sec:conclusions}

We employed population synthesis models of intermediate-mass stars that include the effects of rotation, binary star companions, and measurement uncertainty to study populations of $\delta$\,Sct pulsators. We have verified the accuracy of our synthetic population by simulating the properties of clusters. We chose the recently-characterised Cep--Her Complex, which consists of multiple stellar associations and a star cluster, and which contains many young $\delta$\,Sct pulsators. We showed that the observed distribution of asteroseismic large separations, $\Delta\nu$, is well reproduced by our models. We also solved the mystery of why Cep--Her appeared to have an excess of slow rotators: binary stars, which are preferentially slow rotators, needed to be included in population models.

We have calculated a new semi-empirical instability strip that accounts for the mapping between stars' true properties and their observed properties, where the latter are dependent on the observational inclination angle, presence of unresolved companions, and measurement uncertainty. The instability strip boundaries therefore lie at temperatures hotter than those of the existing theoretical instability strips, and the true strip is slightly narrower. Consequently, the average $\delta$\,Sct star is more massive and younger than previously thought (1.81\,M$_{\odot}$ and 566\,Myr, rather than the 1.71\,M$_{\odot}$ and 657\,Myr implied by previous work). The new instability strip is calibrated against the pulsator-fraction calculations made with 2000 $\delta$\,Sct stars and 12000 non-pulsators from \textit{Kepler}, hence is robust against outliers. The two-sigma bounds on mass that contain 95\% of $\delta$\,Sct stars are 1.50--2.30\,M$_{\odot}$.

The purity of the instability strip is a long-standing question. Field stars have a pulsator fraction peaking at 50-70\% but our simulations expect >90\% of stars in the middle of the instability strip to pulsate, even after observational scatter is accounted for. Hence, some 20--40\% of field stars\footnote{Excludes stars younger than $\lesssim150$\,Myr, which are often not field stars but lie in associations.} that lie inside the instability strip must not pulsate.

We have investigated the period--luminosity relation of our synthetic population and tested hypotheses concerning the recently discovered second ridge. We showed that the second ridge cannot be caused by binary stars. Instead, we found that the observed distribution is well-matched by mixture models consisting of some stars pulsating in the fundamental radial mode ($n=1$), and others in different overtones, predominantly $n=5$ and 6. The appropriate mixture is sample-dependent: samples containing younger and/or less luminous stars consist of fewer fundamental-mode pulsators, perhaps predominantly driven by turbulent pressure. There is also a (real) spread of stars located to the right of the fundamental mode ridge, which we propose are rapid rotators seen equator-on, but in real data one must be careful of outliers whose modes could be misidentified.

We have also predicted the overall $\Delta\nu$ distribution for $\delta$\,Sct stars that exhibit regular patterns in their pulsation modes. Such stars will typically be late pre-MS or young MS stars, and their $\Delta\nu$ distribution has a mean and standard deviation of 6.46 and 0.68\,d$^{-1}$, respectively. We compared our simulated distribution to observations from machine learning and found excellent agreement at high $\Delta\nu \gtrsim 6$\,d$^{-1}$, but at lower values certain features of the observed distribution appear spurious.
Our simulated distribution, or a sample-specific counterpart when studying young associations, could be used as a prior when measuring $\Delta\nu$ via the \'echelle method, to combat bias. 

Finally, in an appendix we have compared the \'echelle and autocorrelation methods for measuring $\Delta\nu$, and presented arguments that the autocorrelation method is less reliable. The autocorrelation method remains useful for its speed and ease of use, but we recommend that asteroseismic large spacings be measured via the \'echelle method, which pulsation models have shown to be accurate.

\section*{Acknowledgements}

SJM was supported by the Australian Research Council through Future Fellowship FT210100485. We thank Marc-Antoine Dupret, Tim Bedding, Tom Love, and Daniel Reese for discussions, along with the USyd--UniSQ $\delta$\,Sct Research Group. We thank the anonymous referee for their helpful suggestions.


\section*{Data Availability}

Data used in this work have had their original source cited. There are no tables of new data in this work.



\bibliographystyle{mnras}
\interlinepenalty=10000
\bibliography{sjm_bibliography} 


\appendix

\section{Contrasting Autocorrelation with the \'Echelle method for measuring $\Delta\nu$}
\label{app:autocorr}

A common technique for finding and parametrising regular spacings in the amplitude spectra of $\delta$\,Sct stars is the autocorrelation method. For instance, \citet{reeseetal2017} used it to measure $\Delta\nu$ and rotational splittings in models of rapidly rotating stars. \citet{beddingetal2020} also used the autocorrelation method to find regular spacings in $\delta$\,Sct stars observed by TESS, and subsequently as an initial guess for $\Delta\nu$ before applying the \'echelle method, although this detail didn't make it into their methods section.
\citet{pamosortegaetal2022,pamosortegaetal2023} used autocorrelation, along with similar methods such as the histogram of frequency differences \citep{handleretal1997} and taking a Fourier transform of the amplitude spectrum \citep{garciahernandezetal2009,garciahernandezetal2013}, to determine $\Delta\nu$ for $\delta$\,Sct stars in $\alpha$\,Per, Trumpler\,10 and Praesepe. Aside from the \'echelle, the three techniques above were summarised by \citet{ramon-ballestaetal2021}.

Since the discovery that the \'echelles of many $\delta$\,Sct stars show a dipole mode ridge and a radial mode ridge, it is apparent that autocorrelation methods often have their strongest peak at a lag of $\Delta\nu/2$ \citep[][and references therein]{beddingetal2020}. This suggests that autocorrelation should be used alongside a prior -- a range of plausible $\Delta\nu$ values -- to avoid measuring the wrong peak. Even then, autocorrelation should not be used blindly. The existence of g\:modes, harmonics in the amplitude spectra of eclipsing binaries, tidally induced pulsations, combination frequencies, or other phenomena that can give rise to regularities in the amplitude spectrum, mean that the autocorrelation method can be misled. To mitigate some of these, one can limit the frequency range upon which the autocorrelation is applied \citep{ramon-ballestaetal2021}, e.g. to frequencies exceeding that of the fundamental radial mode.

It is important to recognise that the \'echelle method and the autocorrelation method do not measure the same thing. The \'echelle method aims to make the radial ridge vertical between $n=5$ and $n=9$ \citep{murphyetal2023},\footnote{\citet{beddingetal2020} used a less formal definition that averaged the values of $\Delta\nu$ needed to make individual ridges vertical.} and this is the method applied to the models in this work. But outside of that range, the radial ridge curves substantially.\footnote{The dipole ridge also contains small curves over its entire length but is roughly parallel to the radial ridge from $n=5$. Ultimately, the definition of $\Delta\nu$ is the frequency difference between modes of the same degree at consecutive radial orders, and it should be measured such that it does not depend which degree is used.}  Thus, the $\Delta\nu$ value measured by the autocorrelation method may differ from the \'echelle method even for a well-behaved star with clear ridges. A final difference is that the \'echelle diagram (but not autocorrelation) measures $\epsilon$, which is the x-position of the radial mode ridge and which is well known to correlate with stellar parameters, both for solar-like oscillators \citep{whiteetal2011b} and $\delta$\,Sct stars \citep{murphyetal2023}.

\begin{figure}
    \centering
    \includegraphics[width=0.995\linewidth]{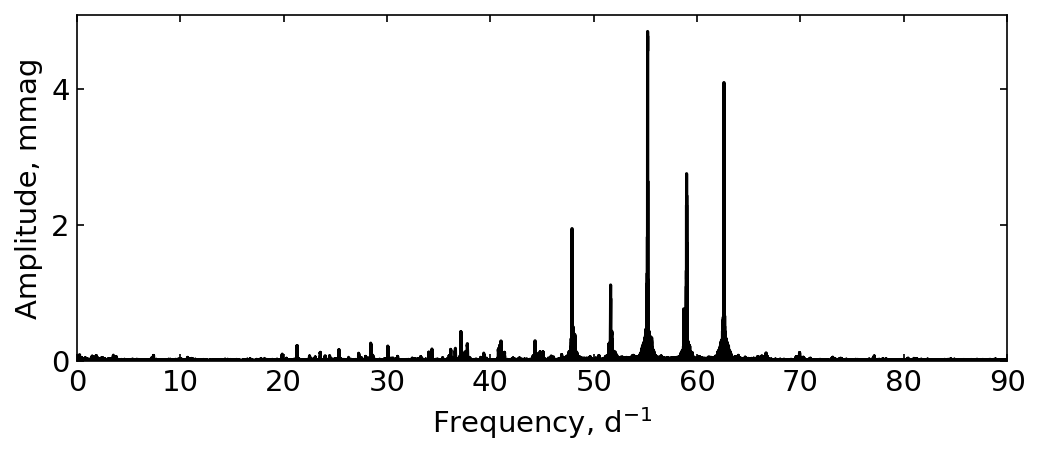}\\
    \includegraphics[width=0.995\linewidth]{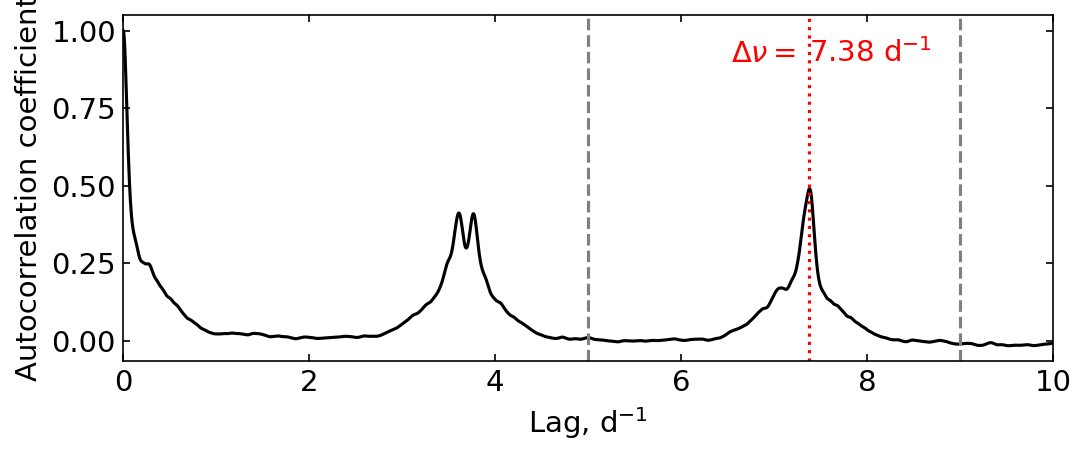}\\
    \includegraphics[width=0.995\linewidth]{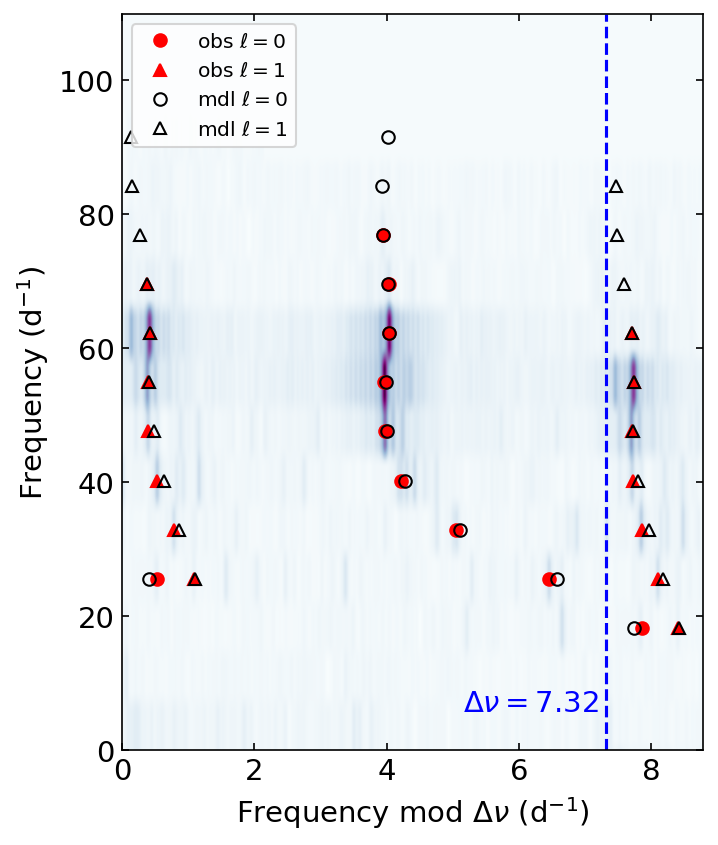}
    \caption{Fourier transform of the TESS 2-min light curve (top), autocorrelation of the amplitude spectrum (middle) and \'echelle diagram (bottom) for TIC\,169464993 in the Cep--Her Complex, as a well-behaved example with two clear ridges. The Fourier amplitudes are convolved with a Gaussian of width 0.05\,d$^{-1}$ before calculation of the autocorrelation and \'echelle. Dashed grey lines in the autocorrelation plot indicate the bounds of the uniform prior on $\Delta\nu$. Note that $\Delta\nu/2$ is also evident, but double-peaked in this case. The value of $\Delta\nu$ from the \'echelle is similar (although not within the standard $\pm$0.02\,d$^{-1}$ uncertainty of the \'echelle method) and indicated with the dashed blue line. The plot area to the right of this line is mirrored from the left-hand side, which is helpful when the dipole ridge crosses the $\Delta\nu$ line, as in Fig.\,\ref{fig:bad_example2}. The shaded background corresponds to Fourier amplitude, and red symbols identify the observed radial and dipole modes. Open black symbols are model frequencies from the model grid in \citet{murphyetal2023}. The x-value of the radial mode ridge is 4/7.32 (or 0.55, hence $\epsilon =1.55$) for this star, but such information is not discernible from the autocorrelation.}
    \label{fig:good_example}
\end{figure}

\begin{figure}
    \centering
    \includegraphics[width=0.995\linewidth]{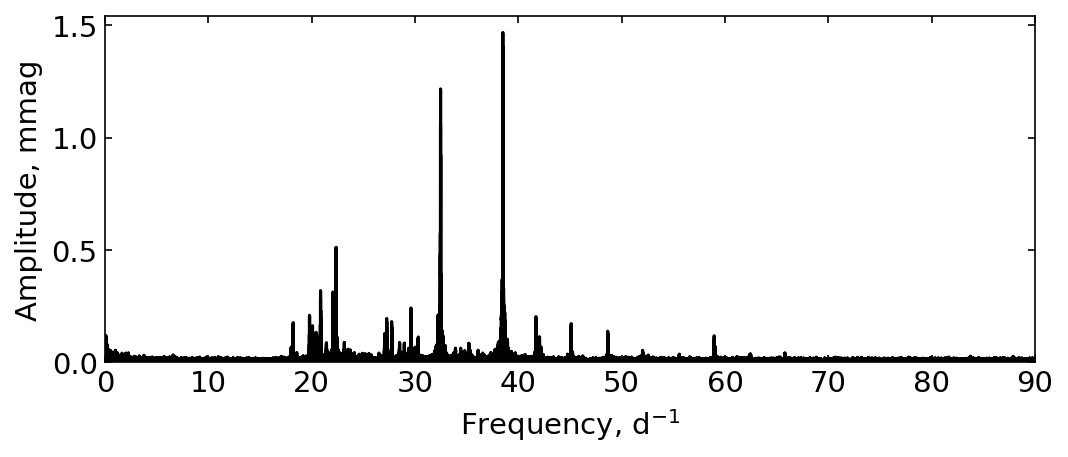}\\
    \includegraphics[width=0.995\linewidth]{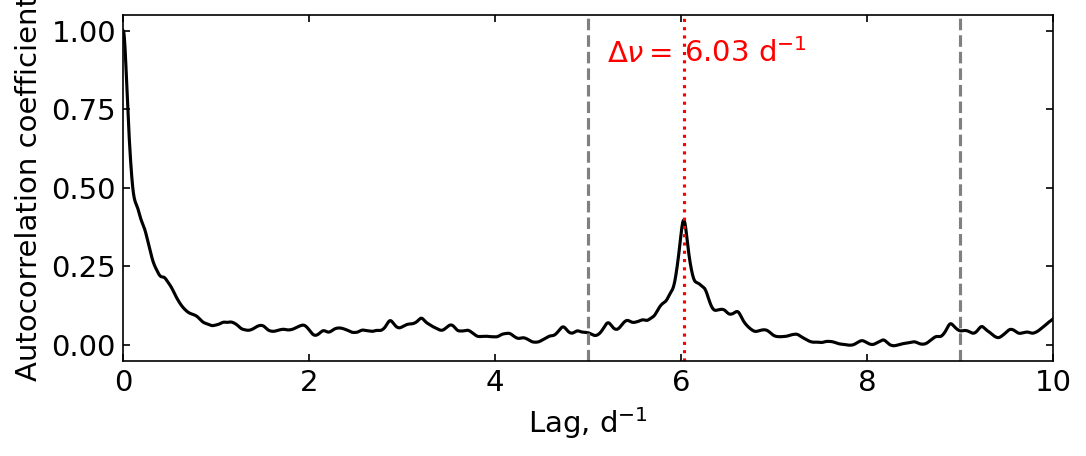}\\
    \includegraphics[width=0.995\linewidth]{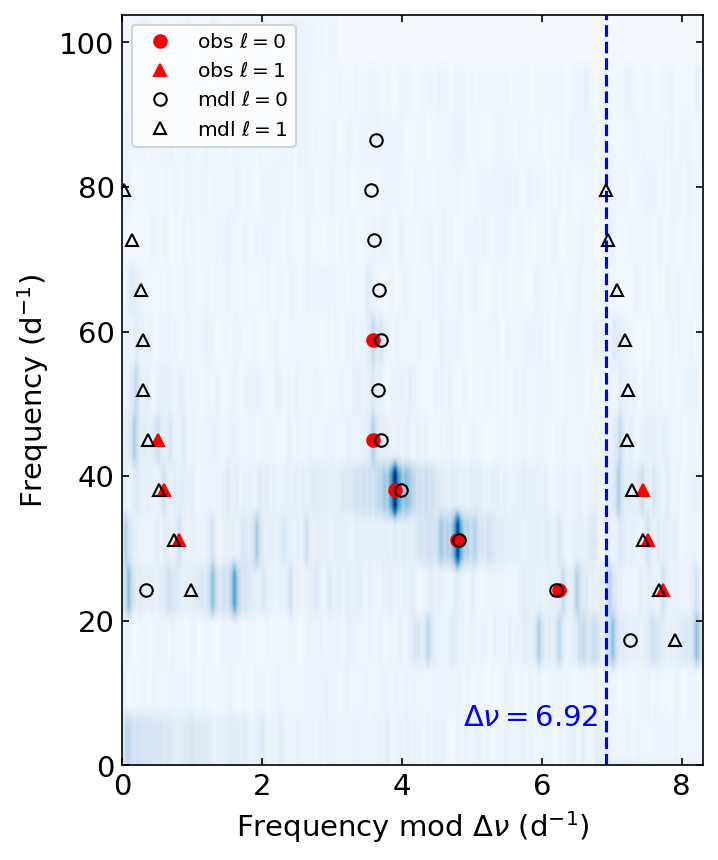}
    \caption{As Fig.\,\ref{fig:good_example}, but for TIC\,295893690 in the Cep--Her Complex.}
    \label{fig:bad_example1}
\end{figure}

\begin{figure}
    \centering
    \includegraphics[width=0.995\linewidth]{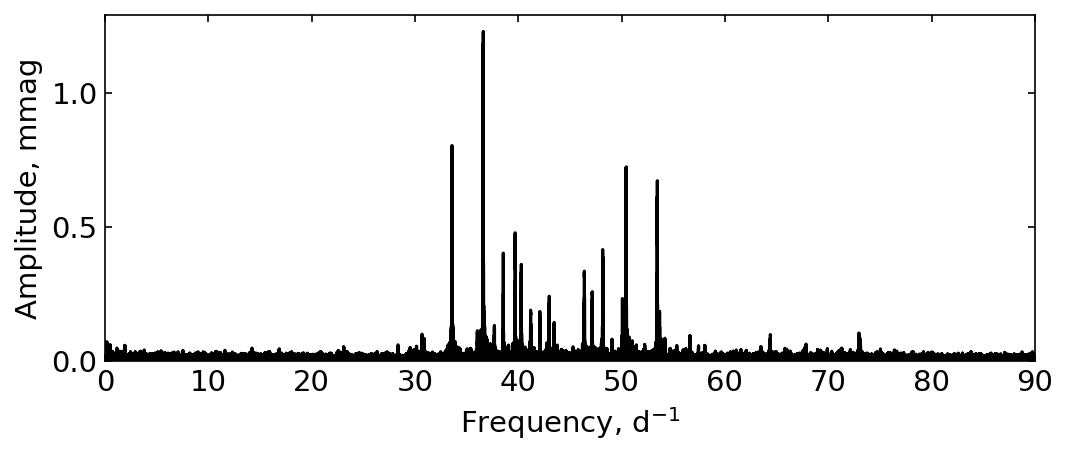}\\
    \includegraphics[width=0.995\linewidth]{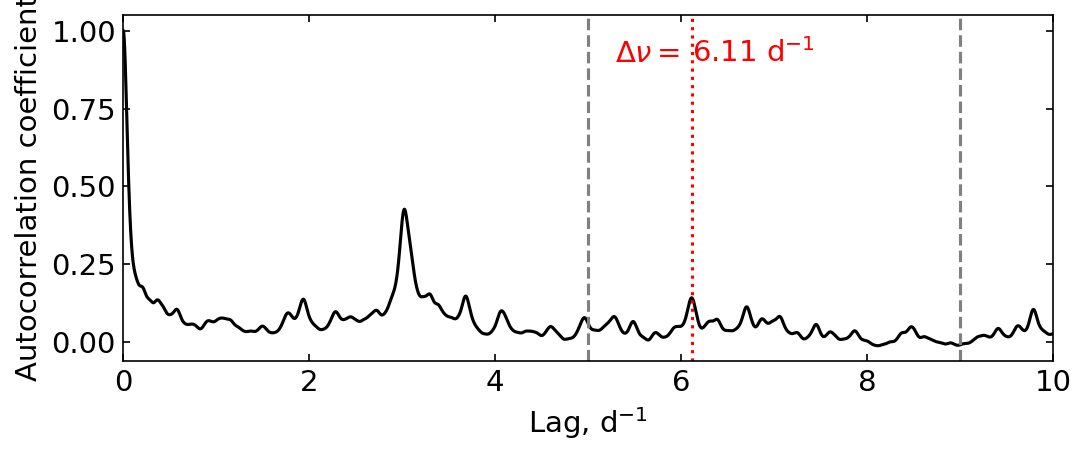}\\
    \includegraphics[width=0.995\linewidth]{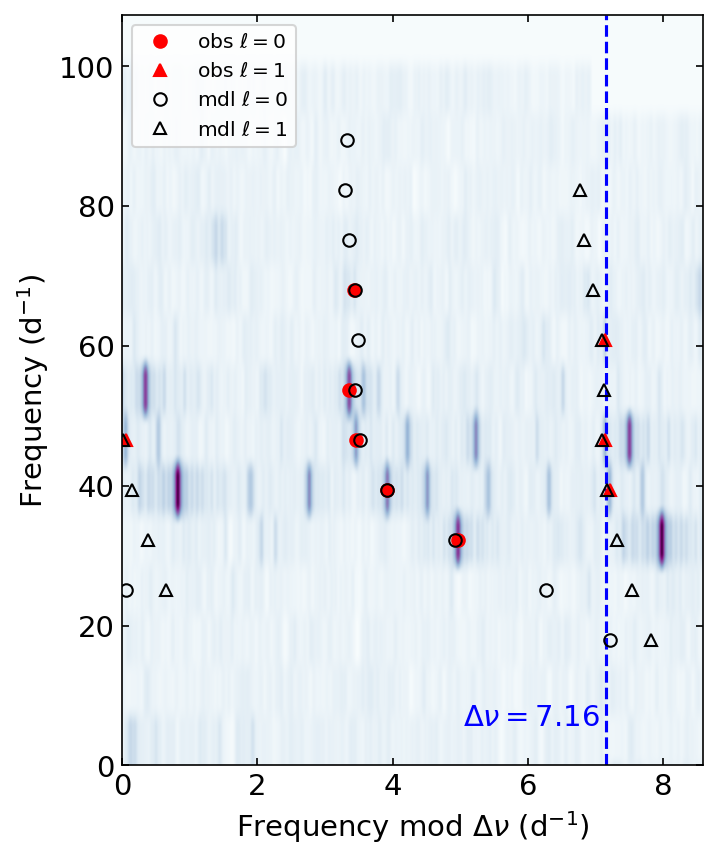}
    \caption{As Fig.\,\ref{fig:good_example}, but for TIC\,171591531 in the Cep--Her Complex.}
    \label{fig:bad_example2}
\end{figure}

In Fig.\,\ref{fig:good_example} we present a well-behaved example where both methods agree. We have used a prior that $\Delta\nu$ should lie in the range 5.0--9.0\,d$^{-1}$ and we only used frequencies above 20\,d$^{-1}$ in the calculation.
Stars with such clear ridges are in the minority, even in young stellar populations like Cep--Her. Figures\,\ref{fig:bad_example1} and \ref{fig:bad_example2} present two mediocre cases where the autocorrelation does not give satisfactory results. Fig.\,\ref{fig:bad_example1} shows TIC\,295893690 from Cep--Her, which has a sharp peak in its autocorrelation function, corresponding to $\Delta\nu=6.03$\,d$^{-1}$, and no peak at $\Delta\nu/2$. The \'echelle, on the other hand, shows a good result at $\Delta\nu=6.92$\,d$^{-1}$ that is well-matched by models. This disagreement arises because many of the observed modes lie at low $n$, where neither the radial ridge nor the dipole ridge are vertical. Moreover, each ridge has a different spacing between its constituent modes at $n<5$, so no single value of $\Delta\nu$ can make both ridges vertical. This is why $\Delta\nu$ is calculated for orders of $n=5$ to $n=9$.

Fig.\,\ref{fig:bad_example2} is a qualitatively different example. Unlike the others, this star does not have just two clear ridges. Rather, three radial orders (three \'echelle rows) contain many peaks. The autocorrelation does not have a dominant peak in the relevant frequency range, but does have a peak that could be interpreted as $\Delta\nu/2$. The strongest peak is at 6.11\,d$^{-1}$ and does not correspond to the $\Delta\nu$ value determined via the \'echelle (and matched with models). 

It is clear from Figures\,\ref{fig:bad_example1} and \ref{fig:bad_example2} that autocorrelation cannot be used blindly, and does not always produce the correct $\Delta\nu$ (i.e. one that corresponds accurately to the square-root of the mean stellar density). It should be noted, of course, that \'echelle diagrams are not free of ambiguity either, and some user expertise is required in more subtle cases such as Figures\,\ref{fig:bad_example1} and \ref{fig:bad_example2}. For instance, a common pitfall is to select $\Delta\nu/2$ rather than $\Delta\nu$ because there are two main ridges (radial, dipole) per \'echelle order. Nonetheless, the manual process of optimising $\Delta\nu$ in an \'echelle diagram does reduce the risk of gross error compared to autocorrelation techniques.

\section{Tabulated period--luminosity relations}
\label{app:PLRs}

For convenience and side-by-side comparison, $P$--$L$ and $P$--$L$--$T_{\rm eff}$ relations are given in Table\:\ref{tab:PLRs}.

\begin{table}
    \centering
    \caption{Period--luminosity and period--luminosity--temperature relations of the form $\log L/{\rm L}_{\odot} = a \log P/{\rm d} + b \log T_{\rm eff}/{\rm K} + c$. The data sets in the first column are those described in Table\:\ref{tab:popsynth_data}.}
    \begin{tabular}{lccr}
\toprule
Data set & $a$ & $b$ & \multicolumn{1}{c}{$c$}  \\
\midrule
\multicolumn{4}{c}{period--luminosity}\\
\midrule
{\it original}	&$	1.278	$&	--	&$	2.694	$\\
{\it model}	&$	1.180	$&	--	&$	2.515	$\\
{\it inclined}	&$	1.176	$&	--	&$	2.510	$\\
{\it binary}	&$	1.162	$&	--	&$	2.527	$\\
{\it observed}	&$	1.160	$&	--	&$	2.526	$\\
\midrule
\multicolumn{4}{c}{period--luminosity--temperature}\\
\midrule
{\it original}	&$	1.521	$&$	4.790	$&$	-15.752	$\\
{\it model}	&$	1.520	$&$	4.787	$&$	-15.741	$\\
{\it inclined}	&$	1.524	$&$	4.844	$&$	-15.958	$\\
{\it binary}	&$	1.424	$&$	3.625	$&$	-11.279	$\\
{\it observed}	&$	1.349	$&$	2.573	$&$	-7.271	$\\
\bottomrule
    \end{tabular}
    \label{tab:PLRs}
\end{table}

\bsp	
\label{lastpage}
\end{document}